\documentclass[journal]{style/IEEEtran}
\usepackage{cite}
\usepackage[colorinlistoftodos,prependcaption,textsize=small]{todonotes}
\usepackage{color}
\usepackage{amsmath,amssymb,graphicx,booktabs,url}
\usepackage[utf8]{inputenc}
\usepackage{microtype}
\usepackage{graphicx}
\usepackage{subfigure}
\usepackage{booktabs} 
\usepackage{amsmath}
\usepackage{amssymb}
\usepackage{stackengine}
\usepackage{tikz}
\usetikzlibrary{shapes.multipart}
\usepackage{multirow}
\usepackage{flushend}

\begin{document}

\newcommand\barbelow[1]{\stackunder[1.2pt]{$#1$}{\rule{.8ex}{.075ex}}}
\newcommand{\A}{\barbelow{{A}}}
\newcommand{\R}{\barbelow{{R}}}
\newcommand{\Ah}{\barbelow{{\hat{A}}}}
\newcommand{\Z}{\barbelow{{Z}}}
\newcommand{\Y}{\barbelow{{Y}}}
\newcommand{\sa}{\barbelow{{a}}}
\newcommand{\s}{\barbelow{{a}}}
\newcommand{\sr}{\barbelow{{r}}}
\newcommand{\sy}{\barbelow{{r}}}
\newcommand{\ah}{\barbelow{{\hat{a}}}}
\newcommand{\sh}{\barbelow{{\hat{a}}}}
\newcommand{\gh}{\barbelow{{\hat{g}}}}
\newcommand{\alfa}{\barbelow{{\alpha}}}
\newcommand{\bet}{\barbelow{{\beta}}}
\newcommand{\e}{\barbelow{{e}}}
\newcommand{\One}{\barbelow{{1}}}
\newcommand{\Zero}{\barbelow{{0}}}
\newcommand{\E}{\mathbb{E}}
\newcommand{\rhoAAh}{\rho \left(\A , \Ah \right)}
\newcommand{\rhoAAhR}{\rho \left(\A , \Ah\left( \R \right) \right)}
\newcommand{\rhoaahr}{\rho \left(\sa , \ah\left( \sr \right) \right)}
\newcommand{\rhoaa}{\rho \left(\sa , \ah \right)}
\newcommand{\far}{ f_{\A , \R} \left( \sa , \sr \right)}
\newcommand{\fagr}{ f_{\A \lvert \R} \left( \sa \lvert \sr \right)}
\newcommand{\fr}{ f_{\R} \left( \sr \right)}
\newcommand{\I}{ \Gamma \left( \sr \right) }
\newcommand{\AhR}{ \Ah\left( \R \right) }
\newcommand{\ahr}{ \ah\left( \sr \right) }
\newcommand{\Hidem}{ \barbelow{\barbelow{{H}}}}
\newcommand{\Agr}{ \A \lvert \sr} 
\newcommand{\Agri}{ {A \lvert r}_i}

\tikzset{%
	do path picture/.style={%
		path picture={%
			\pgfpointdiff{\pgfpointanchor{path picture bounding box}{south west}}%
			{\pgfpointanchor{path picture bounding box}{north east}}%
			\pgfgetlastxy\x\y%
			\tikzset{x=\x/2,y=\y/2}%
			#1
		}
	},
	cross/.style={do path picture={    
			\draw [line cap=round] (-1,-1) -- (1,1) (-1,1) -- (1,-1);
	}},
}

%
\title{On the Relationship Between Short-Time Objective Intelligibility and Short-Time Spectral-Amplitude Mean-Square Error for Speech Enhancement}

\author{\IEEEauthorblockN{Morten Kolbæk,
	Zheng-Hua Tan, Senior Member, IEEE, and
	Jesper Jensen}

\thanks{Manuscript received month day, year; revised month day, year; accepted month day, year. Date of publication month day, year; date of current version Month day, year. This research was partly funded by the Oticon Foundation. The associate editor coordinating the review of this manuscript and approving it for publication was xxyyzz xxyyzz.}%

\thanks{M. Kolbæk and Z.-H. Tan are with the Department of Electronic Systems, Aalborg University,
	Aalborg 9220, Denmark (e-mail: mok@es.aau.dk; zt@es.aau.dk).}%

\thanks{J. Jensen is with the Department of Electronic Systems, Aalborg University, Aalborg 9220, Denmark, and also with Oticon A/S, Smørum 2765, Denmark (e-mail: jje@es.aau.dk; jesj@oticon.com).}%

\thanks{Digital Object Identifier 00.0000/TASLP.2018.0000000}	
}

\maketitle

\begin{abstract}
The majority of deep neural network\,(DNN) based speech enhancement algorithms rely on the mean-square error\,(MSE) criterion of short-time spectral amplitudes\,(STSA), which has no apparent link to human perception, e.g. speech intelligibility. 
Short-Time Objective Intelligibility\,(STOI), a popular state-of-the-art speech intelligibility estimator, on the other hand, relies on linear correlation of speech temporal envelopes. This raises the question if a DNN training criterion based on envelope linear correlation\,(ELC) can lead to improved speech intelligibility performance of DNN based speech enhancement algorithms compared to algorithms based on the STSA-MSE criterion.    	
In this paper we derive that, under certain general conditions, the STSA-MSE and ELC criteria are practically equivalent, and we provide empirical data to support our theoretical results.   
Furthermore, our experimental findings suggest that the standard STSA minimum-MSE estimator is near optimal, if the objective is to enhance noisy speech in a manner which is optimal with respect to the STOI speech intelligibility estimator.
\end{abstract}


\section{Introduction}
\label{sec:intro}
\IEEEPARstart{D}{espite} the recent success of deep neural network\,(DNN) based speech enhancement algorithms \cite{erdogan_deep_2017,wang_deep_2017,wang_supervised_2017,kim_bitwise_2017,fakoor_reinforcement_2017}, it is yet unknown if these algorithms are optimal in terms of aspects related to human auditory perception, e.g. speech intelligibility, since existing algorithms do not directly optimize criteria designed with human auditory perception in mind.

Many current state-of-the-art DNN based speech enhancement algorithms use a mean squared error\,(MSE) training criterion \cite{chen_large-scale_2016,healy_algorithm_2017,kolbaek_speech_2017} on short-time spectral amplitudes\,(STSA).
This, however, might not be the optimal training criterion if the target is the human auditory system, and improvement in speech intelligibility or speech quality is the desired objective.  

It is well known that the frequency sensitivity of the human auditory system is non-linear ( e.g. \cite{schnupp_auditory_2011,moore_introduction_2013}) and, as a consequence, is often approximated in digital signal processing algorithms using e.g. a Gammatone filter bank \cite{patterson_complex_1992} or a one-third octave band filter bank \cite{taal_algorithm_2011}. It is also well known that preservation of modulation frequencies in the range 4-20 Hz are critical for speech intelligibility \cite{elliott_modulation_2009,schnupp_auditory_2011,drullman_effect_1994}.  
Therefore, it is natural to believe that, if prior knowledge about the human auditory system is incorporated into a speech enhancement algorithm, improvements in speech intelligibility or speech quality can be achieved \cite{lim_enhancement_1979}.

Indeed, numerous works exist that attempt to incorporate such knowledge (e.g.  \cite{healy_algorithm_2015,loizou_speech_2005,hendriks_dft-domain_2013,lightburn_sobm_2015,han_perceptual_2016,shivakumar_perception_2016,koizumi_dnn-based_2017,kolbaek_monaural_2018-1,zhao_perceptually_2018,zhang_training_2018,fu_end--end_2018} and references therein).
In \cite{healy_algorithm_2015} a transform-domain method based on a Gammatone filter bank was used, which incorporates a non-linear frequency resolution mimicking that of the human auditory system. 
In \cite{loizou_speech_2005} different perceptually motivated cost functions were used to derive STSA clean speech spectrum estimators in order to emphasize spectral peak information, account for auditory masking or penalize spectral over-attenuation. 
In \cite{han_perceptual_2016,shivakumar_perception_2016} similar goals were pursued, but instead of using classical statistically-based models, DNNs were used.
Finally, in \cite{koizumi_dnn-based_2017} a deep reinforcement learning technique was used to reward solutions that achieved a large score in terms of perceptual evaluation of speech quality\,(PESQ) \cite{rix_perceptual_2001}, a commonly used speech quality estimator.   

Although the works in e.g. \cite{loizou_speech_2005,healy_algorithm_2015,shivakumar_perception_2016,koizumi_dnn-based_2017} include knowledge about the human auditory system the techniques are not designed specifically to maximize speech intelligibility. 
While speech processing methods that improve speech intelligibility would be of vital importance for applications such as mobile communications, or hearing assistive devices, only very little research has been performed to understand if DNN-based speech enhancement systems can help improve speech intelligibility.
Very recent work \cite{kolbaek_monaural_2018-1,zhao_perceptually_2018,zhang_training_2018,fu_end--end_2018} has investigated if DNNs trained to maximize a state-of-the-art speech intelligibility estimator are capable of improving speech intelligibility as measured by the estimator \cite{kolbaek_monaural_2018-1,zhao_perceptually_2018,zhang_training_2018} or human listeners \cite{fu_end--end_2018}.     
Specifically, DNNs were trained to maximize the short-time objective intelligibility\,(STOI) \cite{taal_algorithm_2011} estimator and were then compared, in terms of STOI, with DNNs trained to minimize the classical STSA-MSE criterion. Surprisingly, although all DNNs improved STOI, the DNNs trained to maximize STOI showed none or only very modest improvements in STOI compared to the DNNs trained with the classical STSA-MSE criterion \cite{kolbaek_monaural_2018-1,zhao_perceptually_2018,zhang_training_2018,fu_end--end_2018}.  

The STOI speech intelligibility estimator has proven to be able to quite accurately predict the intelligibility of noisy/processed speech in a large range of acoustic scenarios, including speech processed by mobile communication devices \cite{jorgensen_speech_2015}, ideal time-frequency weighted noisy speech \cite{taal_algorithm_2011}, noisy speech enhanced by single-microphone time-frequency weighting-based speech enhancement systems \cite{jensen_algorithm_2016,taal_algorithm_2011,jensen_speech_2014}, and speech processed by hearing assistive devices such as cochlear implants \cite{falk_objective_2015}. STOI has also been shown to be robust to variations in language types, including Danish \cite{taal_algorithm_2011}, Dutch \cite{jensen_speech_2014}, and Mandarin \cite{xia_evaluation_2012}. 
Finally, recent studies e.g. \cite{healy_algorithm_2017,chen_large-scale_2016} also show a good correspondence between STOI predictions of noisy speech enhanced by DNN-based speech enhancement systems, and speech intelligibility.  
As a consequence, STOI is currently the, perhaps, most commonly used speech intelligibility estimator for objectively evaluating the performance of speech enhancement systems \cite{healy_algorithm_2015,chen_large-scale_2016,healy_algorithm_2017,kolbaek_speech_2017}.  
Therefore, it is natural to believe that gains in speech intelligibility, as estimated by STOI, can be achieved by utilizing an optimality criterion based on STOI as opposed to the classical criterion based on STSA-MSE.  

In this paper we study the potential gain in speech intelligibility that can be achieved, if a DNN is designed to perform optimally with respect to the STOI speech intelligibility estimator. 
We derive that, under certain general conditions, maximizing an approximate-STOI criterion is equivalent to minimizing a STSA-MSE criterion. 
Furthermore, we present empirical data using simulation studies with DNNs applied to noisy speech signals, that support our theoretical results. 
Finally, we show theoretically under which conditions the equality between the approximate-STOI criterion and the STSA-MSE criterion holds for practical systems. 
Our results are in line with recent empirical work and might explain the somewhat surprising result in \cite{kolbaek_monaural_2018-1,zhao_perceptually_2018,zhang_training_2018,fu_end--end_2018}, where none or only very modest improvements in STOI were achieved with STOI optimal DNNs compared to MSE optimal DNNs.

\section{STFT-domain based Speech Enhancement}\label{sec:secse}
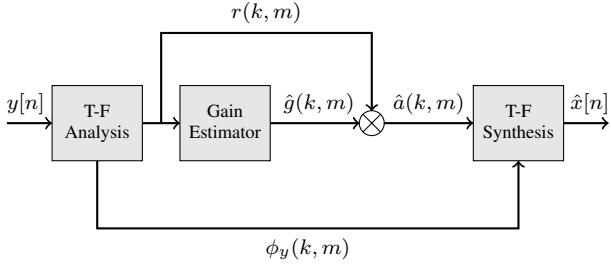
\begin{figure}
	\centering
	\vspace{2mm}
	\begin{tikzpicture}[baseline=(current bounding box.north)]
	\def\x{0.2}
	
	\draw [thick, ->] (0,2) -- (0.6,2);
	\node[above] at (0.25,2) {\footnotesize $y[n]$};
	\draw [thick, ->] (1.8,2) -- (2.3,2);
	\node[above] at (3.45,3.2) {\footnotesize $r(k,m)$};
	\draw [thick, ->]  (2.05,2) -- (2.05,3.2) -- (4.85,3.2) -- (4.85,2.16) ;
	\draw [thick, ->]  (1.2,1.5) -- (1.2,0.6) -- (6.8,0.6) -- (6.8,1.5) ;
	\node[below] at (4,0.6) {\footnotesize $\phi_y(k,m)$};
	\draw [thick, ->] (3.5,2) -- (4.69,2);
	\node[above] at (4.15,2.00) {\footnotesize $\hat{g}(k,m)$};
	
	\node [draw,circle,cross,minimum width=0.2 ] at (4.85,2){}; 
	
	\draw [thick, ->] (5.01,2) -- (6.2,2);
	\node[above] at (5.61,2.00) {\footnotesize $\hat{a}(k,m)$};
	
	\draw [thick, ->] (7.2,2) -- (8,2);
	\node[above] at (7.75,2) {\footnotesize $\hat{x}[n]$};

	\filldraw[fill=gray!20!white, draw=black] (0.6 , 1.5) rectangle (1.8 , 2.5) ;
	\node at (1.2,2) {\scriptsize {\begin{tabular}{c} T-F \\ Analysis \end{tabular}}};
	
	\filldraw[fill=gray!20!white, draw=black] (2.3 , 1.5) rectangle (3.5 , 2.5) ;
	\node at (2.9,2) {\scriptsize {\begin{tabular}{c} Gain \\ Estimator \end{tabular}}};
	
	\filldraw[fill=gray!20!white, draw=black] (6.2 , 1.5) rectangle (7.4 , 2.5) ;
	\node at (6.8,2) {\scriptsize {\begin{tabular}{c} T-F \\ Synthesis \end{tabular}}};

	\end{tikzpicture}
	\caption{Classical gain-based speech enhancement system. The noisy time-domain signal $y[n] = x[n] + v[n]$ is first decomposed into a time-frequency\,(T-F) representation $r(k,m)$ for time-frame $m$ and frequency index $k$. An estimator, e.g. a DNN, estimates a gain $\hat{g}(k,m)$ that is applied to the noisy short-term magnitude spectrum $r(k,m)$ to arrive at an enhanced signal magnitude  $\hat{a}(k,m) = \hat{g}(k,m)r(k,m)$. Finally, the enhanced time-domain signal $\hat{x}[n]$ is obtained from a T-F synthesis stage using the phase of the noisy signal $\phi_y(k,m)$.}
    \label{fig:sefig}
\end{figure}

Fig.\;\ref{fig:sefig} shows a block-diagram of a classical gain-based speech enhancement system \cite{hendriks_dft-domain_2013,loizou_speech_2013}.
%
%
Let $x[n]$ be the $n$th sample of the clean time-domain speech signal and let a noisy observation $y[n]$ be given by
\begin{equation}
y[n] = x[n] + v[n], 
\label{eq111}
\end{equation}
where $v[n]$ is a sample of additive noise.   
Furthermore, let $a(k,m)$ and $r(k,m)$,  $k = 1,\dots, \frac{K}{2}+1$,  $m=1,\dots M, $ denote the single-sided magnitude spectra of the $K$-point short-time discrete Fourier transform (STFT) of $x[n]$ and $y[n]$, respectively, where $M$ is the number of STFT frames.  
Also, let $\hat{a}(k,m)$ denote an estimate of $a(k,m)$ obtained as $\hat{a}(k,m) = \hat{g}(k,m)r(k,m)$. 
Here, $\hat{g}(k,m)$ is a scalar gain factor applied to the magnitude spectrum of the noisy speech $r(k,m)$ to arrive at an estimate $\hat{a}(k,m)$ of the clean speech magnitude spectrum $a(k,m)$. It is the goal of many STFT-based speech enhancement systems to find appropriate values for $\hat{g}(k,m)$ based on the available noisy signal $y[n]$.    
The gain factor $\hat{g}(k,m)$ is typically estimated using either statistical model-based methods such as classical STSA minimum mean-square error\,(MMSE) estimators \cite{ephraim_speech_1984}, \cite{hendriks_dft-domain_2013,loizou_speech_2013}, or machine learning based techniques such as Gaussian mixture models \cite{kim_algorithm_2009}, support vector machines \cite{han_classification_2012}, or, more recently, DNNs \cite{healy_algorithm_2015,chen_large-scale_2016,healy_algorithm_2017,kolbaek_speech_2017}. 
For reconstructing the enhanced speech signal in the time domain, it is common practice to append the short-time phase spectrum of the noisy signal to the estimated short-time magnitude spectrum and then use the overlap-and-add technique \cite{allen_short_1977}, \cite{loizou_speech_2013}.

\section{Short-Time Objective Intelligibility (STOI)}\label{sec:secstoi}
In the following, we shortly review the STOI intelligibility estimator \cite{taal_algorithm_2011}. For further details we refer to \cite{taal_algorithm_2011}. 
Let the $j$th one-third octave band clean-speech amplitude, for time-frame $m$, be defined as 
\begin{equation}
a_j( m ) = \sqrt{\sum_{k=k_1(j)}^{k_2(j)} a(k,m)^2},
\label{eq22}
\end{equation}
where $k_1(j)$ and $k_2(j)$ denote the first and last STFT bin index, respectively, of the $j$th one-third octave band.
Furthermore, let a short-time temporal envelope vector that spans time-frames $m-N+1, \dots, m$, for the clean speech signal be defined as 
\begin{equation}
\s_{j,m} = [ a_j( m-N+1 ), \; a_j( m-N+2 ), \dots , a_j( m ) ]^T
\label{eq222}
\end{equation}
In a similar manner we define  $\sh_{j,m}$ and $\sy_{j,m}$ for the enhanced speech signal and the noisy observation, respectively. 

The parameter $N$ defines the length of the temporal envelope and for STOI $N=30$%
\footnote{With $N=30$, STOI is sensitive to temporal modulations of $2.6$ Hz and higher, which are frequencies important for speech intelligibility \cite{taal_algorithm_2011}.}, which for the STFT settings used in this study, as well as in \cite{taal_algorithm_2011}, corresponds to approximately $384$ ms.
Finally, the STOI speech intelligibility estimator for a pair of short-time temporal envelope vectors can then be approximated by the sample envelope linear correlation\,(ELC) between the clean and enhanced envelope vectors $\s_{j,m}$ and $\sh_{j,m}$ given as 
\begin{equation}
\mathcal{L} ( \s_{j,m},\sh_{j,m}) = \frac{\left(\s_{j,m} - \mu_{\s_{j,m}}\right)^T  \left(\sh_{j,m} - \mu_{\sh_{j,m}}\right)}{ \left\lVert \s_{j,m} - \mu_{\s_{j,m}} \right\rVert  \; \left\lVert\sh_{j,m} - \mu_{\sh_{j,m}} \right\rVert },
\label{eq:stoicost}
\end{equation}
where $\left\lVert \cdot \right\rVert$ denotes the Euclidean $\ell^2$-norm and $\mu_{\s_{j,m}}$ and $\mu_{\sh_{j,m}}$ denote the sample means of $\s_{j,m}$ and $\sh_{j,m}$, respectively. 
Note that Eq.\;\eqref{eq:stoicost} is an approximation, since the clipping and normalization steps otherwise used in STOI, have been omitted. This has empirically been found not to have any significant effect on intelligibility prediction performance in most cases  \cite{taal_matching_2012,lightburn_sobm_2015,jensen_algorithm_2016,andersen_predicting_2016}. 
Furthermore, since the normalization step is applied for the entire vector $\sh_{j,m}$, the normalization procedure itself does not influence the final STOI score. Also, as clipping only occurs for time-frequency units for which the signal-to-distortion ratio (see Eq. (4) in \cite{taal_algorithm_2011}) is below $-15$ dB, clipping only occurs for a minority of the envelope vectors and approximating STOI with ELC is well valid, or even exact, in most cases, when evaluating speech signals at practical SNRs.

From $\mathcal{L} ( \s_{j,m},\sh_{j,m})$, the final STOI score for an entire speech signal is then defined as \cite{taal_algorithm_2011} the scalar, $-1 \leq d \leq 1$,    
\begin{equation}
d = \frac{1}{J(M-N+1)} \sum_{j=1}^{J} \sum_{m=N}^{M} \mathcal{L} ( \s_{j,m},\sh_{j,m}),
\label{eq:stoisum}
\end{equation}
where $J$ is the number of one-third octave bands and $M-N+1$ is the total number of short-time temporal envelope vectors.

Similarly to \cite{taal_algorithm_2011}, we use $J=15$ with a center frequency of the first one-third octave band at 150 Hz and the last at approximately 3.8 kHz to ensure a frequency range that covers the majority of the spectral information of human speech.
The STOI score in general has been shown to often have high correlation with listening tests involving human test subjects, i.e. the higher numerical value of Eq.\;\eqref{eq:stoisum}, the more intelligible is the speech signal.    

Since STOI, as approximated by Eq.\;\eqref{eq:stoisum}, is a sum of ELC values as given by Eq.\;\eqref{eq:stoicost}, maximizing Eq.\;\eqref{eq:stoicost} will also maximize the overall STOI score in Eq.\;\eqref{eq:stoisum}. 
As a consequence, in order to find an estimate $\hat{x}[n]$ of $x[n]$ so that STOI is maximized, one can focus on finding optimal estimates of the individual short-time temporal envelope vectors $\s_{j,m}$.      
Therefore, we define $\sh_{j,m} = \text{diag}(\gh_{j,m})\sy_{j,m}$ as the short-time temporal one-third octave band envelope vector of the enhanced speech signal, where $\gh_{j,m}$ is an estimated gain vector and $\text{diag}(\gh_{j,m})$ is a diagonal matrix with the elements of $\gh_{j,m}$ on the main diagonal.

\section{Envelope Linear Correlation Estimator}\label{sec:melcest}
We now introduce the approximate-STOI criterion in a stochastic context and derive the speech envelope estimator that maximizes it. We denote this estimator as the maximum mean envelope linear correlation\,(MMELC) estimator. 
Let $A_j(m)$ and $R_j(m)$ denote random variables representing a clean and a noisy, respectively, one-third octave band magnitude, for band $j$ and time frame $m$. Furthermore, let   
\begin{equation}
\A_j(m) = \left[ A_j(m-N+1), \, \dots \, A_j(m) \right]
\label{eq1}
\end{equation}
and
\begin{equation}
\R_j(m) = \left[ R_j(m-N+1), \, \dots \, R_j(m) \right]
\label{eq2}
\end{equation}
be the stack of these random variables in random envelope vectors. 
Finally, in a similar manner, let 
\begin{equation}
\Ah_j(m) = \left[ \hat{A}_j(m-N+1), \, \dots \, \hat{A}_j(m) \right], 
\label{eq3}
\end{equation}
be a random envelope vector representing an estimate of $\A_j(m)$.  
Now, the contribution of $\Ah_j(m)$ to speech intelligibility may be approximated
as the ELC between the envelope vectors $\A_j(m)$ and $\Ah_j(m)$. In the following, the indices $j$ and $m$ are omitted for convenience. 
Let $\One$ denote a vector of ones, and let $\barbelow{\mu}_{\A} = \frac{1}{N} \One^T \A \One$ be a vector, whose entries equal the sample mean of the entries in $\A$.
Let  $\barbelow{\mu}_{\Ah}$ be defined in a similar manner. 
Finally, let the ELC between $\A$ and $\Ah$, which is a random variable, be defined as
\begin{equation}
\rhoAAh \triangleq \frac{ \Big( \A - \barbelow{\mu}_{\A} \Big)^T  \Big( \Ah - \barbelow{\mu}_{\Ah} \Big) }{ \Big\lVert \A - \barbelow{\mu}_{\A} \Big\rVert \Big\lVert \Ah - \barbelow{\mu}_{\Ah} \Big\rVert },
\label{eq4}
\end{equation} 
and the expected ELC as
\begin{equation} 
\begin{split}
\Omega_{ELC}	& = \E_{\A,\R}  \left[ \rhoAAh \right]  \\
& = \int \int \rhoaa \far  d \sa \, d \sr \\
& = \int \underbrace{ \int \rhoaa \fagr \, d \sa \; }_{\I} \fr \, d \sr .
\end{split}
\label{eq5}
\end{equation}    
Here, $\ah$ is related to $\sr$ via a deterministic map, e.g. a DNN, and $\far$ denotes the joint probability density function\,(PDF) of clean and noisy/processed one-third octave band envelope vectors. Furthermore, $\fagr$ and $\fr$ denote a conditional and marginal PDF, respectively.

An optimal estimator can be found by minimizing the Bayes risk \cite{kay_fundamentals_2010,loizou_speech_2013}, which is equivalent to maximizing Eq.\;\eqref{eq5}, hence arriving at the MMELC estimator, which we denote as $\ah_{MMELC}$.  
To do so, observe that for a particular noisy observation $\sr$ maximizing $\I$ maximizes Eq.\;\eqref{eq5}, since $\fr \geq 0 \; \forall \; \sr$. In other words, our goal is to maximize $\I$ for each and every $\sr$. Hence, for a particular observation, $\sr$, the MMELC estimate is given by
\begin{equation} 
\begin{split}
\ah_{MMELC} & = \arg\max_{\ah} \int \rhoaa \fagr \, d \sa \\
& = \arg\max_{\ah} \int \frac{ \big( \sa - \barbelow{\mu}_{\sa} \big)^T  \big( \ah - \barbelow{\mu}_{\ah} \big) }{ \big\lVert \sa - \barbelow{\mu}_{\sa} \big\rVert \big\lVert \ah - \barbelow{\mu}_{\ah} \big\rVert } \fagr \, d \sa  \\
& = \arg\max_{\ah} \underbrace{ \int \frac{ \big( \sa - \barbelow{\mu}_{\sa} \big)^T }{ \big\lVert \sa - \barbelow{\mu}_{\sa} \big\rVert }   \fagr \, d \sa }_{\E_{\A | \sr } \left[  \e(\A)^T \right] }   \underbrace{ \frac{\big( \ah - \barbelow{\mu}_{\ah} \big)}{\big\lVert \ah - \barbelow{\mu}_{\ah} \big\rVert} }_{\e(\ah)}  \\
& = \arg\max_{\ah} \; \E_{\A | \sr } \left[ \e(\A)^T \right] \e(\ah),  \\
\end{split}
\label{eq6}
\end{equation}     
where $\e(\cdot)$ is a function that normalizes its vector argument to zero sample mean and unit norm 
and where we used that for a given noisy observation $\sr$, $\ah$ is deterministic. 
Note that the solution to Eq.\;\eqref{eq6} is non-unique. For one given solution, say $\ah^\ast$, any affine transformation, $\delta \ah^\ast + \gamma \One \; \forall \; \delta,\gamma \, \in \mathcal{R} $, is also a solution, because any such transformation is undone by $\e(\cdot)$.
Hence, in the following we focus on finding one such particular solution, namely the zero sample mean, unit norm solution, i.e. the vector $\e(\ah)$ that maximizes the inner product with the vector $\E_{\A | \sr } \left[ \e(\A| \sr) \right]$. 
To do so, let $\alfa = \E_{\A | \sr } \left[ \e(\A| \sr) \right]$, and let $\e(\ah^\ast)$ denote the zero sample mean, unit norm vector that maximizes Eq.\;\eqref{eq6}. 
Then, using the method of Lagrange multipliers, it can be shown (see Appendix\;\ref{sec:lagran}) that the MMELC estimator is given by
\begin{equation} 
\begin{split}
\ah_{MMELC} &=  \e(\ah^\ast) \\
& = \frac{\big( \alfa - \barbelow{\mu}_{\alfa} \big)}{\big\lVert \alfa - \barbelow{\mu}_{\alfa} \big\rVert} \\
& = \frac{\alfa}{\Vert \alfa \Vert}, 
\end{split}
\label{eq7}
\end{equation}  
which is nothing more than the vector $\alfa$, normalized to unit norm. 
The fact that $\barbelow{\mu}_{\alfa} = \frac{1}{N}  \One^T \alpha \One = \Zero$ follows from Eq.\;\eqref{eq6}, where it is seen that $\alfa = \E_{\A | \sr } \left[ \e(\A| \sr) \right]$ is an expectation over vectors $ (\sa - \barbelow{\mu}_{\sa})  \big\lVert \sa - \barbelow{\mu}_{\sa} \big\rVert^{-1} $ whose sample mean is zero. 
By interpreting the expectation as an infinite linear combination of such vectors, it follows that $\barbelow{\mu}_{\alfa} = \Zero$.

\section{Relation to STSA-MMSE Estimators}\label{secrelation}
We now show that the MMELC estimator, Eq.\;\eqref{eq7}, is asymptotically equivalent to the one-third octave band STSA-MMSE estimator for large envelope lengths, i.e. as $N \to \infty$.
The STSA-MSE (e.g. \cite{ephraim_speech_1984}) is defined as
\begin{equation} 
\begin{split}
\Omega_{MSE}	& = \E_{\A,\R}  \left[ \left(\A-\Ah \right)^2 \right].  \\
\end{split}
\label{eq8}
\end{equation}  
It can be shown (e.g. \cite{ephraim_speech_1984,hendriks_dft-domain_2013,loizou_speech_2013}) that the optimal Bayesian estimator with respect to Eq.\;\eqref{eq8}, is the STSA-MMSE estimator given by the conditional mean defined as 
\begin{equation} 
\begin{split}
\ah_{MMSE} & =  \int \sa \; \fagr \, d \sa  \\
& = \E_{\A | \sr }  \left[ \A | \sr \right]. \\
\end{split}
\label{eq9}
\end{equation}  
To show that $\ah_{MMELC}$ is asymptotically equivalent to $\ah_{MMSE}$, let us introduce the idempotent, symmetric matrix
\begin{equation} 
\Hidem =  \barbelow{\barbelow{I}}_N - \frac{1}{N} \One\One^T,
\label{eq10}
\end{equation} 
where $\barbelow{\barbelow{I}}_N$ denotes the $N$-dimensional identity matrix. We can then rewrite the vector $\alfa$ as
\begin{equation} 
\begin{split}
\alfa & = \int \frac{ \big( \sa - \barbelow{\mu}_{\sa} \big) }{ \big\lVert \ah - \barbelow{\mu}_{\ah} \big\rVert }   \fagr \, d \sa   \\
& =  \int \frac{  \Hidem \sa }{ \big\lVert \Hidem \sa \big\rVert }   \fagr \, d \sa   \\
& = \E_{\A | \sr } \left[ \frac{  \Hidem \A | \sr }{ \big\lVert \Hidem \A | \sr \big\rVert }  \right] \\
& = \E_{\A | \sr } \left[ \frac{  \Z }{ \big\lVert \Z \big\rVert }  \right], \\
\end{split}
\label{eq11}
\end{equation} 
where $\A | \sr $ is a random vector, and we introduced the notation $\Z \triangleq \Hidem \A | \sr$.
We now employ the following conditional independence assumption
\begin{equation} 
\fagr = \prod_{j=1}^{N} f_{A_j | R_j = r_j} (a_j | r_j).
\label{eqz9}
\end{equation} 
This is a standard assumption in the area of speech enhancement, when operating in the STFT domain and has been the underlying assumption of a very large number of speech enhancement methods (see e.g. \cite{ephraim_speech_1984,ephraim_speech_1985,erkelens_minimum_2007,hendriks_dft-domain_2013,loizou_speech_2013} and references therein). 
The conditional independence assumption is, for example, valid, when speech and noise STFT coefficients may be assumed statistically independent across time and frequency and mutually independent \cite{mcaulay_speech_1980,ephraim_speech_1984,loizou_speech_2013}.

Using Kolmogorovs strong law of large numbers  \cite[pp. 67-68]{sen_large_1994} and the conditional independence assumption, it can be shown (see Appendix\;\ref{sec:zindepedent}) that asymptotically, as $N \to \infty$, the expectation in Eq.\;\eqref{eq11} factorizes as   
\begin{equation} 
\begin{split}
\lim_{N\to\infty} \;\; \alfa &= \lim\limits_{N \to \infty} \E_{\A | \sr } \left[ \frac{  1 }{ \big\lVert \Z \big\rVert }  \right]     \E_{\A | \sr }\left[ \Z  \right]. \\
\end{split}
\label{eq14}
\end{equation} 
Combining this result with Eq.\;\eqref{eq7} leads to
\begin{equation} 
\begin{split}
\lim_{N\to\infty} \ah_{MMELC} &=  \lim_{N\to\infty} \frac{  \alfa }{\big\lVert \alfa \big\rVert}\\
&= \lim_{N\to\infty}  \frac{   \E_{\A | \sr } \left[ \frac{  1 }{ \lVert \Z \rVert }  \right]     \E_{\A | \sr } \left[ \Z  \right]  }{ \Big\lVert \E_{\A | \sr } \left[ \frac{  1 }{ \lVert \Z \rVert }  \right]     \E_{\A | \sr } \left[ \Z  \right] \Big \rVert }\\
&=  \lim_{N\to\infty} \frac{   \E_{\A | \sr } \left[ \frac{  1 }{ \lVert \Z \rVert }  \right]     \E_{\A | \sr } \left[ \Z  \right]  }{ \E_{\A | \sr } \left[ \frac{  1 }{ \lVert \Z \rVert }  \right] \Big\lVert    \E_{\A | \sr } \left[ \Z  \right]   \Big \rVert   }\\
&= \lim_{N\to\infty} \frac{       \E_{\A | \sr } \left[ \Z  \right]  }{ \Big\lVert    \E_{\A | \sr } \left[ \Z  \right]   \Big \rVert   }.\\
\end{split}
\label{eq15}
\end{equation}  
Since Eq.\;\eqref{eq6} is invariant to affine transformations of its input arguments, we can scale  $\ah_{MMELC} $  with the scalar quantity $  \lVert \E_{\A | \sr } \left[  \Z   \right] \rVert  $ in Eq.\;\eqref{eq15} to arrive at 
\begin{equation} 
\begin{split}
\lim_{N\to\infty} \;\; \ah_{MMELC} &=  \E_{\A | \sr } \left[ \Z  \right].\\
\end{split}
\label{eq16}
\end{equation}  
Finally, as $N \to \infty $, the MMELC estimator  $\ah_{MMELC}$ is given by
\begin{equation} 
\begin{split}
\lim_{N\to\infty} \;\; \ah_{MMELC} &= \;  \E_{\A | \sr } \left[ \Z  \right] \\
&= \E_{\A | \sr } \big[   \Hidem \A | \sr  \big] \\
&=\E_{\A | \sr } \Bigg[ \left(  \barbelow{\barbelow{\mathbf{I}}}_N - \frac{1}{N} \One\One^T \right) \A | \sr  \Bigg] \\
&=\E_{\A | \sr } \bigg[  \A | \sr - \frac{1}{N} \One\One^T  \A | \sr  \bigg] \\
&=\E_{\A | \sr } \big[  \A | \sr \big]  - \frac{1}{N} \One\One^T  \E_{\A | \sr } \big[ \A | \sr  \big]  \\
&=\ah_{MMSE}   - \barbelow{\mu}_{\ah_{MMSE} }.\\
\end{split}
\label{eq17}
\end{equation} 
In words, the MMELC estimator, $\ah_{MMELC}$, is (asymptotically in $N$) an affine transformation of the STSA-MMSE estimator $\ah_{MMSE}$. 
In practice, this means that using the STSA-MMSE estimator leads to the same approximate-STOI criterion value as the estimator, $\ah_{MMELC}$, derived to maximize this criterion.  

In other words, applying the traditional STSA-MMSE estimator leads to maximum speech intelligibility as reflected by the approximate STOI estimator.
\section{Experimental Design}\label{sec:expdes}
We now investigate empirically the relationship between the MMELC estimator in Eq.\;\eqref{eq9} and the STSA-MMSE estimator in Eq.\;\eqref{eq6} using an experimental study.   
As defined in Eq.\;\eqref{eq6}, the MMELC estimator is the vector that maximizes the expectation of the ELC cost function given by Eq.\;\eqref{eq5}. 
This expectation, Eq.\;\eqref{eq5}, is defined via an integral of $\rhoaa$ for various realizations of $\sa$ and $\ah$, and weighted by the joint PDF $\far$. It is however, well known, that the integral may be approximated (arbitrarily well) as a sum of $\rhoaa$ terms, where realizations of $\sa$ and $\ah$ are drawn according to $\far$.      
This is similar to what a DNN approximates during a standard training process, where a gradient based optimization technique is used to minimize the cost on a representative training set \cite{goodfellow_deep_2016}. 
Therefore, training a DNN, e.g. using stochastic gradient ascent, to maximize Eq.\;\eqref{eq:stoicost} may be seen as an approximation of Eq.\;\eqref{eq6}, where the approximation becomes more accurate with increasing training set size. 

From the theoretical results presented in Sec.\;\ref{secrelation}, we would therefore expect that, for some sufficiently large $N$, one would obtain equality in an ELC sense, between a DNN trained to maximize an ELC cost function and one that is trained to minimize the classical STSA-MSE cost function.
To validate this expectation we follow the techniques formalized in Secs.\;\ref{sec:secse} and \ref{sec:secstoi} and train DNNs to estimate gain vectors, $\gh_{j,m}$, that we apply to noisy one-third octave band magnitude envelope signals $\sy_{j,m}$, to arrive at enhanced signals $\sh_{j,m}$. 

In principle, any supervised learning model would be applicable for these experiments but considering the universal function approximation capability of DNNs \cite{hornik_multilayer_1989}, this is our model of choice.
We use short-time temporal one-third octave band envelope vectors, as defined in Eq.\;\eqref{eq222}, and train multiple DNNs, one for each of the $J=15$ one-third octave bands, for various $N$, to investigate if for sufficiently large $N$, DNNs trained with a STSA-MSE cost function approach the ELC values of DNNs trained with a cost function based on ELC. 

We construct two types of enhancement systems, one type is trained using the STSA-MSE cost function, denoted as $\text{ES}_{MSE}$, and one that is trained using the ELC cost function denoted as $\text{ES}_{ELC}$. 
Each of the systems consists of $J=15$ DNNs, each estimating a gain vector $\gh_{j,m}$ for a particular one-third octave band directly from the STFT magnitudes of the noisy signal $r(k,m)$, with the input context given by $k = 1,\dots, \frac{K}{2}+1$,  $m-N+1 \dots, m $. This ensures that all DNNs have access to the same information for a particular value of $N$, as they all receive the same input data.  
Furthermore, we follow common practice (e.g. \cite{healy_algorithm_2015,healy_algorithm_2017,chen_large-scale_2016,kolbaek_monaural_2018-1}) and average overlapping estimated gain values, within a one-third octave band, during enhancement. We found during a preliminary study that this technique consistently lead to slightly larger STOI scores for both types of systems.

To compute the STFT coefficients for all signals we use a 10 kHz sample frequency and a $K=256$ point STFT with a Hann-window size of 256 samples (25.6 ms) and a 128 sample frame shift (12.8 ms). These coefficients are then used to compute one-third octave band envelopes for the clean and noisy signals using Eq.\;\eqref{eq222}.   
\subsection{Noise-free Speech Mixtures}
We have used the Wall Street Journal\,(WSJ0) speech corpus \cite{garofolo_csr-i_1993} as the clean speech data for both the training set, validation set, and test set. Specifically, the noise-free utterances used for training and validation are generated by randomly selecting utterances from 44 male and 47 female speakers from the WSJ0 training set entitled si\_tr\_s. In total 20000 utterances are used for the training set and 2000 are used for the validation set, which adds up to approximately 37 hours of training data and 4 hours of validation data.
For the test set, we have used a similar approach and sampled 1000 utterances among 16 speakers (10 males and 6 females) from the WSJ0 validation set si\_dt\_05 and evaluation set si\_et\_05, which is equivalent to approximately 2 hours of data, see \cite{kolbaek_supplemental_nodate-1} for further details.    
The speakers used in the training and validation sets are different than the speakers used for test, i.e. we test in a speaker independent setting. 
{\color{black} Finally, since WSJ0 utterances primarily include speech active regions we do not apply a VAD. This is motivated by the fact that noise-only regions are irrelevant for STOI, as these are discarded by an ideal VAD in the STOI front-end \cite{taal_algorithm_2011}.}
\subsection{Noise Types}\label{sec:nt}
To simulate a wide variety of sound scenes we have used six different noise types in our experiments: two synthetic noise signals and four natural noise signals, which are real-life recordings of naturally occurring sound scenes.  
For the two synthetic noise signals, we use a stationary speech shaped noise\;(SSN) signal and a highly non-stationary 6-speaker babble\;(BBL) noise.
For the naturally occurring noise signals, we use the street\;(STR), cafeteria\;(CAF), bus\;(BUS), and pedestrian\;(PED) noise signals from the CHiME3 dataset \cite{barker_third_2015}.    
The SSN noise signal is Gaussian white noise, spectrally shaped according to the long-term spectrum of the entire TIMIT speech corpus \cite{garofolo_darpa_1993}. Similarly, the BBL noise signal is constructed by mixing utterances from both genders from TIMIT. 
To ensure that all noise types are equally represented and with unique realizations in the training, validation and test sets, all six noise signals are split into non-overlapping segments such that 40 min.\;is used for training, 5 min.\;is used for validation and another 5 min. is used for test. 
\subsection{Noisy Speech Mixtures}
To construct the noisy speech signals used for training, we follow Eq.\;\eqref{eq111} and combine a noise-free training utterance $x[n]$ with a randomly selected noise sequence $v[n]$, of equal length, from the training noise signal. 
We scale the noise signal $v[n]$, to achieve a certain signal-to-noise ratio\,(SNR), according to the active speech level of $x[n]$ as defined by ITU P.56 \cite{itu_rec._1993}.      
For the training and validation sets, the SNRs are chosen uniformly from $[-5 , 10 ]$ dB to ensure that the intelligibility of the noisy speech mixtures $y[n]$ ranges from degraded to perfectly intelligible.    
\subsection{Model Architecture and Training}
The two types of enhancement systems, $\text{ES}_{ELC}$ and $\text{ES}_{MSE}$, each consist of 15 feed-forward DNNs. The DNNs in the $\text{ES}_{ELC}$ system are trained with the ELC cost function introduced in Eq.\;\eqref{eq:stoicost} and the DNNs in the $\text{ES}_{MSE}$ system are trained using the well-known STSA-MSE cost function given by   
\begin{equation}
\mathcal{J} ( \s,\sh) = \frac{1}{N} \left\lVert \s - \sh \right\lVert^2,
\label{eq:msecost}
\end{equation}
where the subscripts $j$ and $m$ are omitted for convenience.
We train both the $\text{ES}_{ELC}$ and $\text{ES}_{MSE}$ systems with 20000 training utterances and 2000 validation utterances and both data sets have been mixed uniformly with the SSN, BBL, CAF, and STR noise signals, which ensures that each noise type have been mixed with $25\%$ of the utterances in the training and validation sets. 
During test, we evaluate each system with one noise type at a time, i.e. each system is evaluated with 1000 noisy test utterances per noise type, and since BUS and PED are not included in the training and validation sets, these two noise signals serve as unmatched noise types, whereas SSN, BBL, CAF, and STR are matched noise types. This will allow us to study how the ELC optimal DNNs and STSA-MSE optimal DNNs generalize to unmatched noise types.

Each feed-forward DNN consists of three hidden layers with 512 units using ReLU activation functions. The $N$-dimensional output layer uses sigmoid functions which ensures that the output gain $\gh_{j,m}$ is confined between zero and one.
The DNNs are trained using stochastic gradient de-/ascent with the backpropagation technique and batch normalization \cite{goodfellow_deep_2016}. 
The DNNs are trained for a maximum of 200 epochs with a minibatch size of 256 randomly selected short-time temporal one-third octave band envelope vectors.

Since the $\text{ES}_{ELC}$ and $\text{ES}_{MSE}$ systems use different cost functions, they likely have different optimal learning rates. This is easily seen from the gradient norms of the two cost functions. 
It can be shown (details omitted due to space limitations) that the $\ell^2$-norm of the gradient of the ELC cost function in Eq.\;\eqref{eq:stoicost}, with respect to the desired signal vector $\sh$, is given by 
\begin{equation}
\left\lVert \nabla \mathcal{L} ( \s,\sh) \right\rVert =  \frac{ \sqrt{1-\mathcal{L} ( \s,\sh)^2} }{ \left\lVert \sh \right\rVert },
\label{eq:stoigradnorm}
\end{equation}
where the gradient $\nabla \mathcal{L} ( \s,\sh)$ is given by
\begin{equation}
\begin{split}
&\nabla \mathcal{L} \big( \s,\sh \big) = \\
& \left[ \frac{\partial \mathcal{L} \big( \s,\sh \big)}{\partial \hat{{a}}_{1}},
\frac{\partial \mathcal{L} \big( \s,\sh \big) }{\partial \hat{a}_{2}}, \dots,
\frac{\partial \mathcal{L} \big( \s,\sh \big) }{\partial \hat{a}_{N}}
\right]^T,
\label{eq:stoigrad}
\end{split}
\end{equation}
and
\begin{equation}
\begin{split}
&\frac{\partial \mathcal{L} \big( \s,\sh \big)}{\partial \hat{a}_{m}} = \\
&\frac{\mathcal{L} \big( \s,\sh \big)  \big( \s_m - \mu_{\s} \big)    }{  \big(\sh - \mu_{\sh}  \big)^T \big( \s  - \mu_{\s} \big) } -
\frac{\mathcal{L} \big( \s,\sh \big)  \big( \sh_m - \mu_{\sh} \big)    }{  \big(\sh - \mu_{\sh}  \big)^T \big( \sh - \mu_{\sh}  \big)   } . 
\label{eq:stoipart}
\end{split}
\end{equation}
is the partial derivative of $\mathcal{L} ( \s,\sh)$ with respect to entry $m$ of vector $\sh$.
Similarly, the gradient of the STSA-MSE cost function in Eq.\;\eqref{eq:msecost} is given by
\begin{equation}
\begin{split}
&\nabla \mathcal{J} \big( \s,\sh \big) = \frac{2}{N} \left(  \s - \sh \right), 
\label{eq:msegrad}
\end{split}
\end{equation}
such that
\begin{equation}
\begin{split}
\left\lVert  \nabla \mathcal{J} \big( \s,\sh \big) \right\rVert = \frac{2}{N} \left\lVert    \s - \sh   \right\rVert.
\label{eq:msegradnorm}
\end{split}
\end{equation}
Note, since $ \mathcal{L} ( \s,\sh) $ is invariant to the magnitude of $\left\lVert \sh \right\rVert$ (see Eq.\;\eqref{eq:stoicost}), and $\s$ and $N$ are constants during training, the gradient norm of the ELC cost function, Eq.\;\eqref{eq:stoigradnorm}, with respect to $\sh$, is inversely proportional to the gradient norm of the STSA-MSE cost function, Eq.\;\eqref{eq:msegradnorm}.
This suggests that the two cost functions have different optimal learning rates. 
This observation might partly explain why equality with respect to STOI between STOI optimal and STSA-MSE optimal DNNs were achieved in \cite{kolbaek_monaural_2018-1} but not in \cite{zhao_perceptually_2018,zhang_training_2018,fu_end--end_2018}, as \cite{kolbaek_monaural_2018-1} was the only study that explicitly stated that different learning rates for the two cost functions were used. 
In fact, in \cite{zhao_perceptually_2018,zhang_training_2018,fu_end--end_2018} the optimization method Adam \cite{kingma_adam:_2014} was used, and although Adam is an adaptive gradient method, it still has several critical hyper-parameters that can influence convergence \cite{wilson_marginal_2017}.

During a preliminary grid-search using the validation set corrupted with SSN at an SNR of 0 dB and $N=30$, we found learning rates of $0.01$ and $5 \cdot 10^{-5}$ per sample to be optimal for the $\text{ES}_{ELC}$ and $\text{ES}_{MSE}$ systems, respectively.  
During training, the cost on the validation set was evaluated for each epoch and the learning rates were scaled by $0.7$, if the cost increased compared to the cost for the previous epoch. The training was terminated, if the learning rate was below $10^{-10}$. 
We implemented the DNNs using CNTK \cite{agarwal_introduction_2014} and the scripts needed to reproduce the reported results can be found in \cite{kolbaek_supplemental_nodate-1}. 
Note, the goal of these experiments is not to achieve state-of-the-art enhancement performance. In fact, increasing the size of the dataset or DNNs might likely improve performance, although we have not reason to believe it will change the conclusion.

\section{Experimental Results}\label{sec:expres}
To study the relationship between $\text{ES}_{ELC}$ and $\text{ES}_{MSE}$ systems as function of $N$, we have trained multiple systems for various $N$. Specifically, a total of eight $\text{ES}_{ELC}$ systems and eight $\text{ES}_{MSE}$ systems have been trained with $N$ taking the values $N = \{4, 7, 15, 20, 30, 40, 50, 80\}$, which correspond to temporal envelope vectors with durations from approximately 50 to 1000 milliseconds.  
%
%
%
%
%
\begin{figure*}[ht] 
	\centering
	\centerline{\includegraphics[trim={50mm 20mm 50mm 17mm},clip,width=1.0\linewidth]{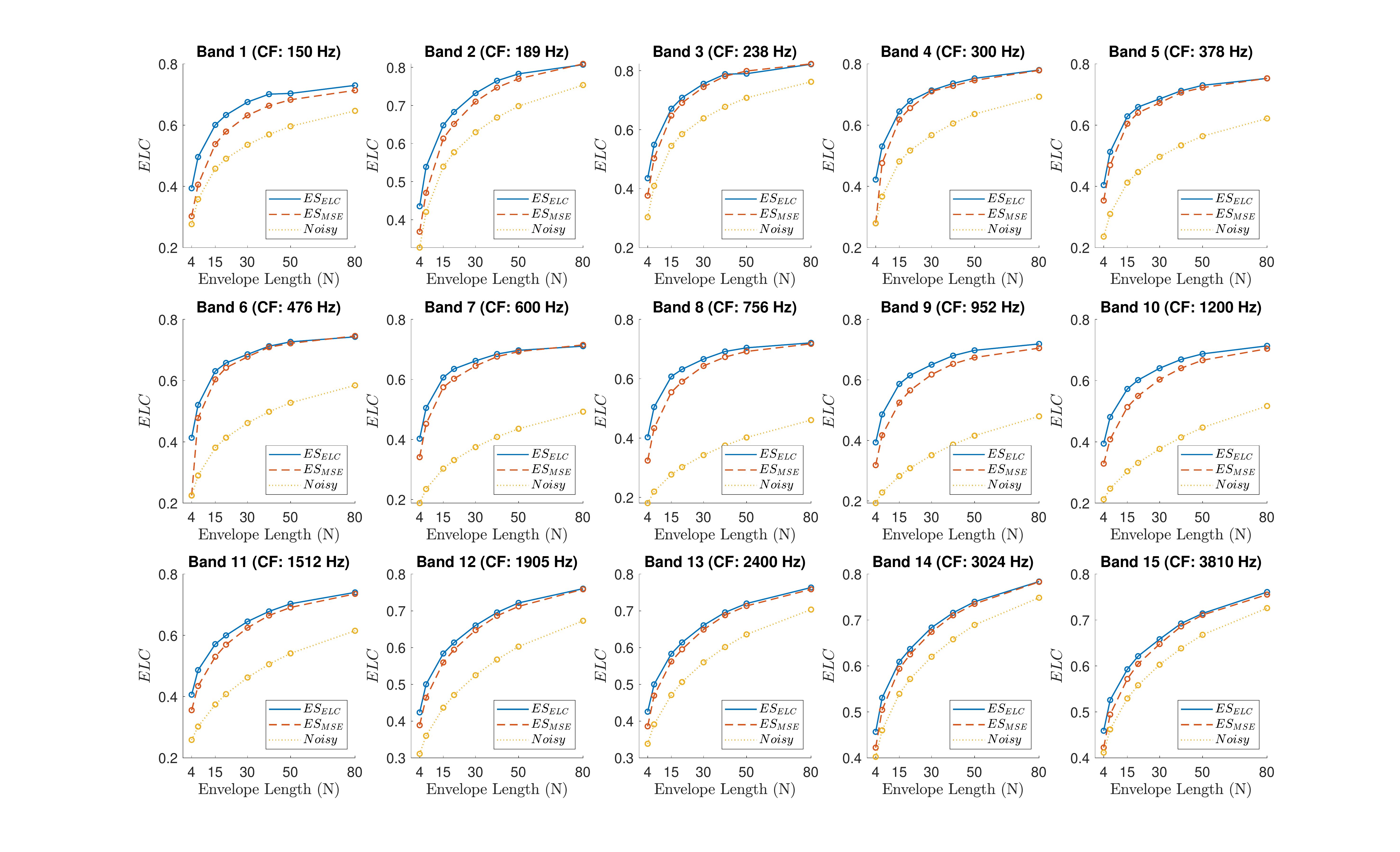}}
	\caption{ELC values for $\text{ES}_{ELC}$ and $\text{ES}_{MSE}$ systems trained using various envelope durations, $N$, and tested with corresponding values of $N$ using speech corrupted with BBL noise at an SNR of 0 dB. Each figure shows one out of $J=15$ one-third octave band DNNs (center frequency\,(CF) shown in parenthesis). It is seen that as $N \to 80$ the difference between the $\text{ES}_{ELC}$ DNNs and $\text{ES}_{MSE}$ DNNs, as measured by ELC, tends to zero. This is in line with the theoretical results of Sec.\;\ref{secrelation}.  }
	\label{fig:perband}
\end{figure*}
\begin{figure*}[ht]
	\centering
	\begin{minipage}{.33\textwidth}
		\centering
		\centerline{\includegraphics[trim={0mm -11mm 10mm 0mm},clip,width=0.9\linewidth]{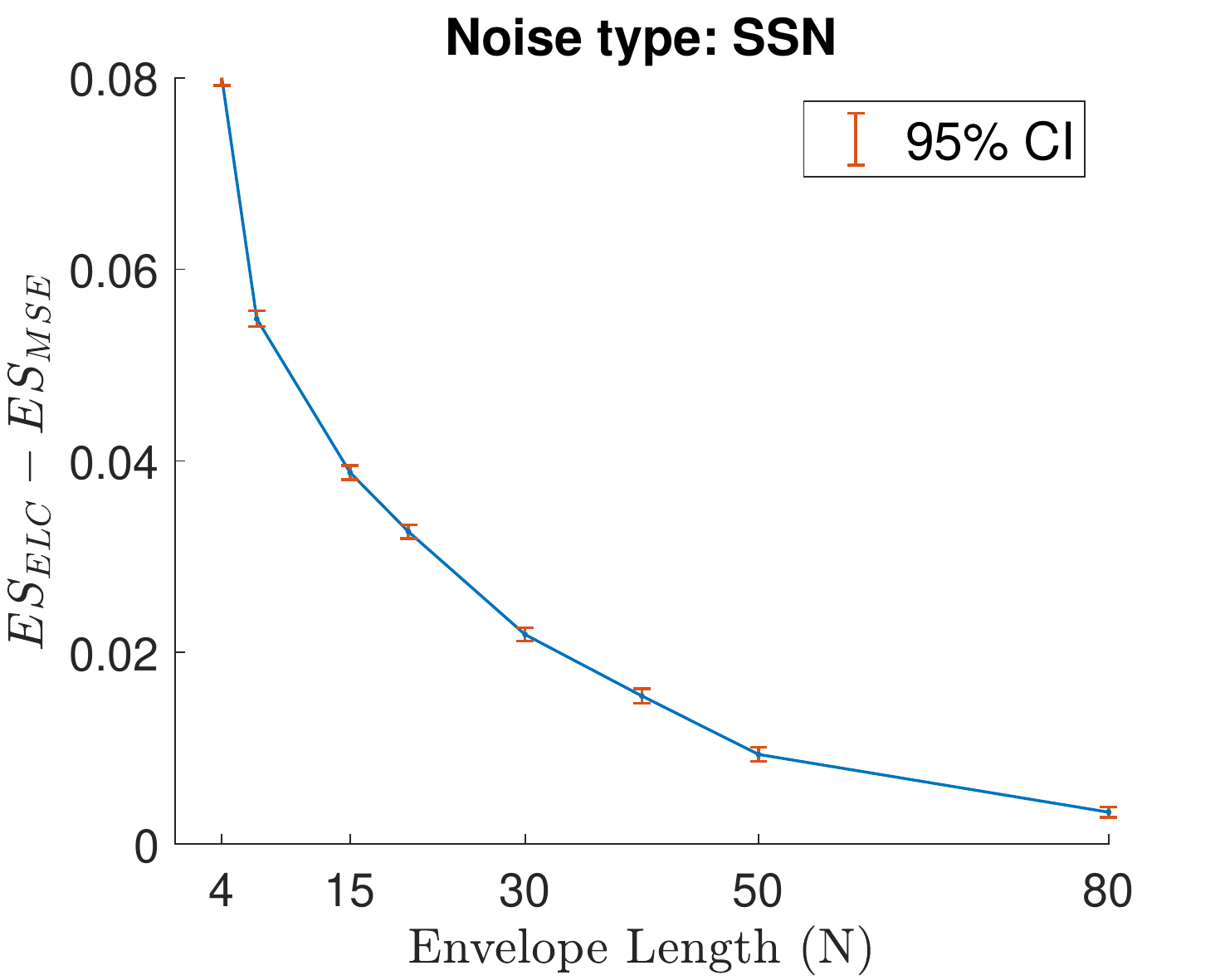}}
		\label{fig:test1}
	\end{minipage}%
	\begin{minipage}{.33\textwidth}
		\centering
		\centerline{\includegraphics[trim={0mm -11mm 10mm 0mm},clip,width=0.9\linewidth]{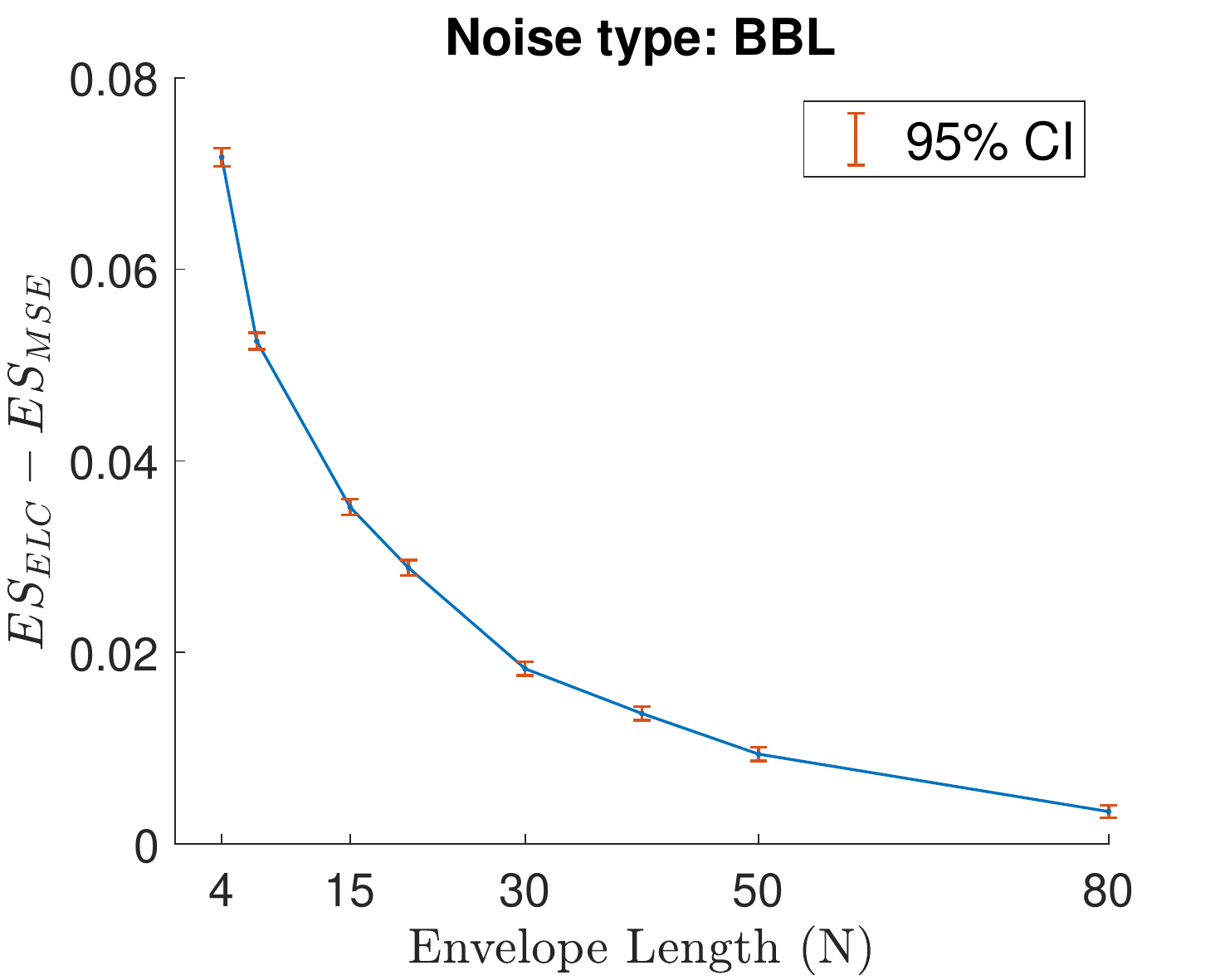}}
		\label{fig:test2}
	\end{minipage}
	\begin{minipage}{.33\textwidth}
		\centering
		\centerline{\includegraphics[trim={0mm -11mm 10mm 0mm},clip,width=0.9\linewidth]{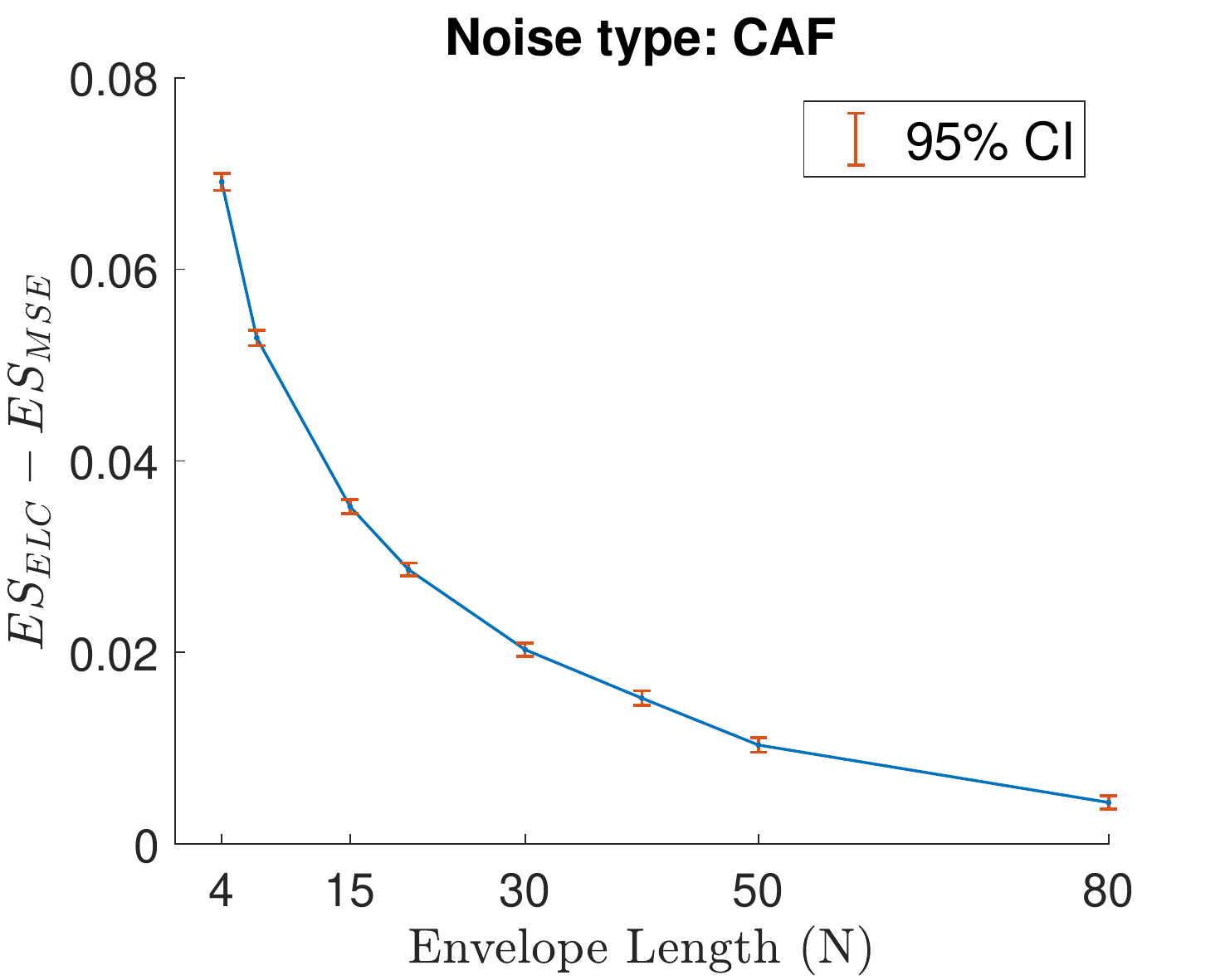}}
		\label{fig:test3}
	\end{minipage}
	\begin{minipage}{.33\textwidth}
		\centering
		\centerline{\includegraphics[trim={0mm 0mm 11mm 0mm},clip,width=0.9\linewidth]{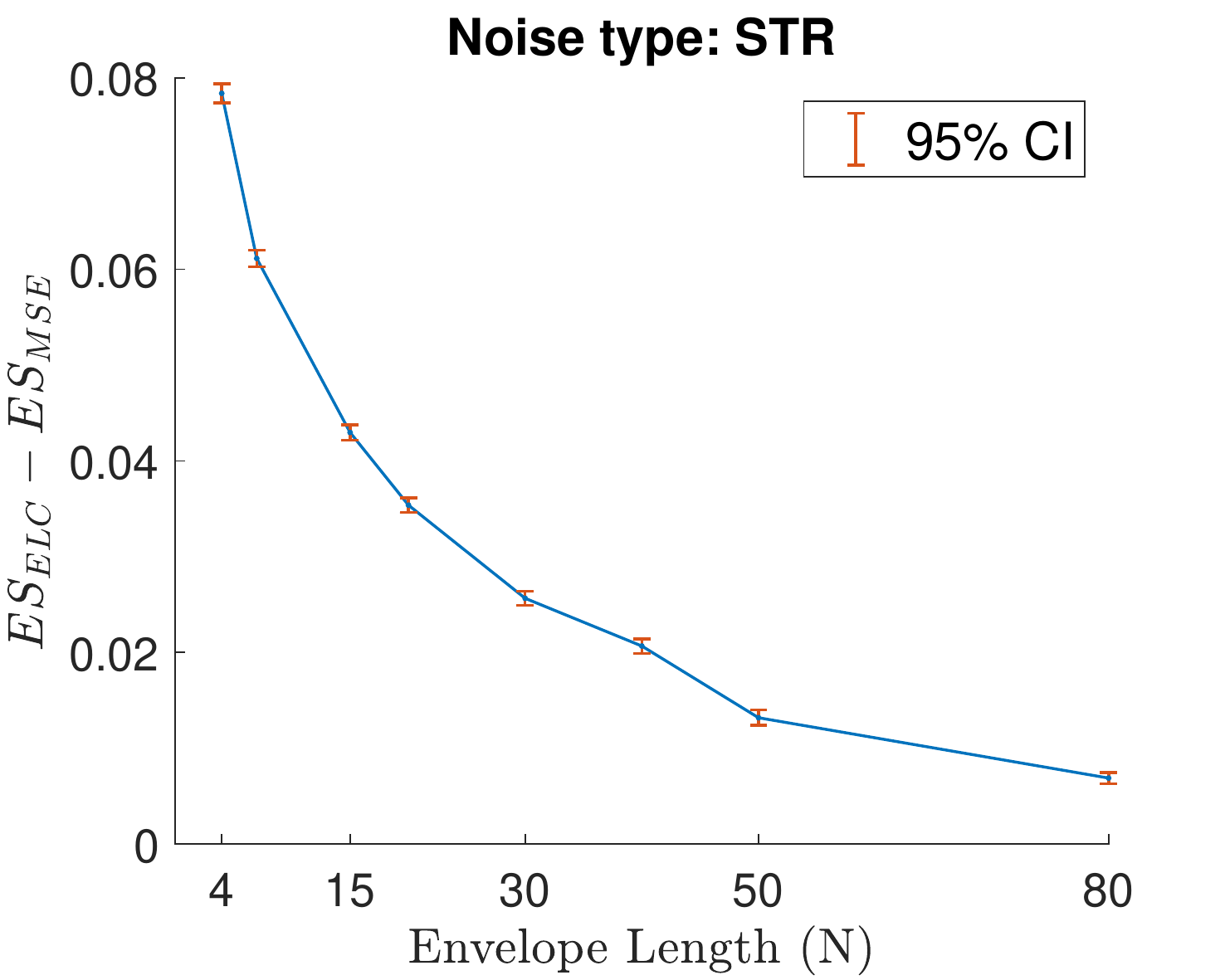}}
		\label{fig:test4}
	\end{minipage}%
	\begin{minipage}{.33\textwidth}
		\centering
		\centerline{\includegraphics[trim={0mm 0mm 11mm 0mm},clip,width=0.9\linewidth]{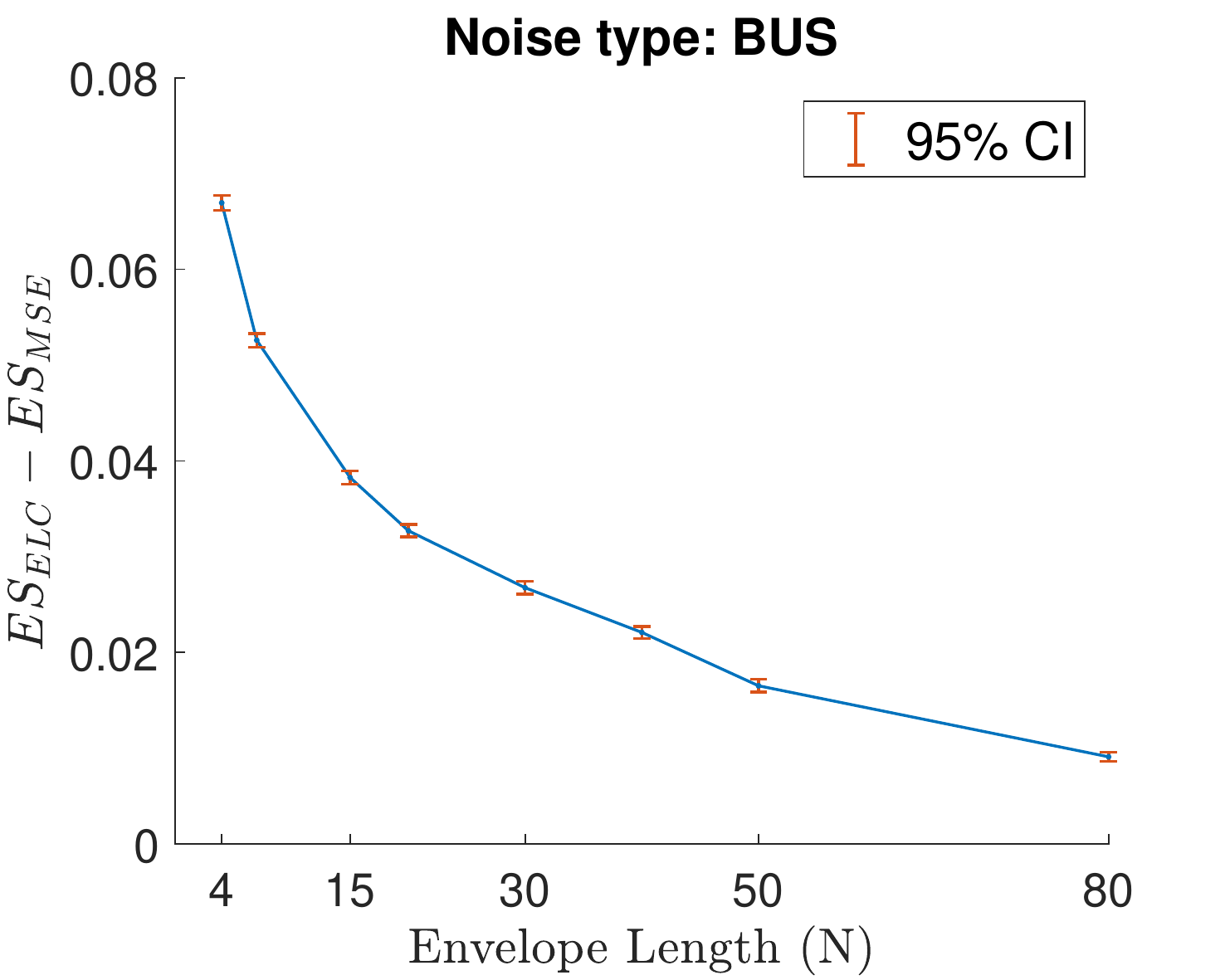}}
		\label{fig:test5}
	\end{minipage}
	\begin{minipage}{.33\textwidth}
		\centering
		\centerline{\includegraphics[trim={0mm 0mm 11mm 0mm},clip,width=0.9\linewidth]{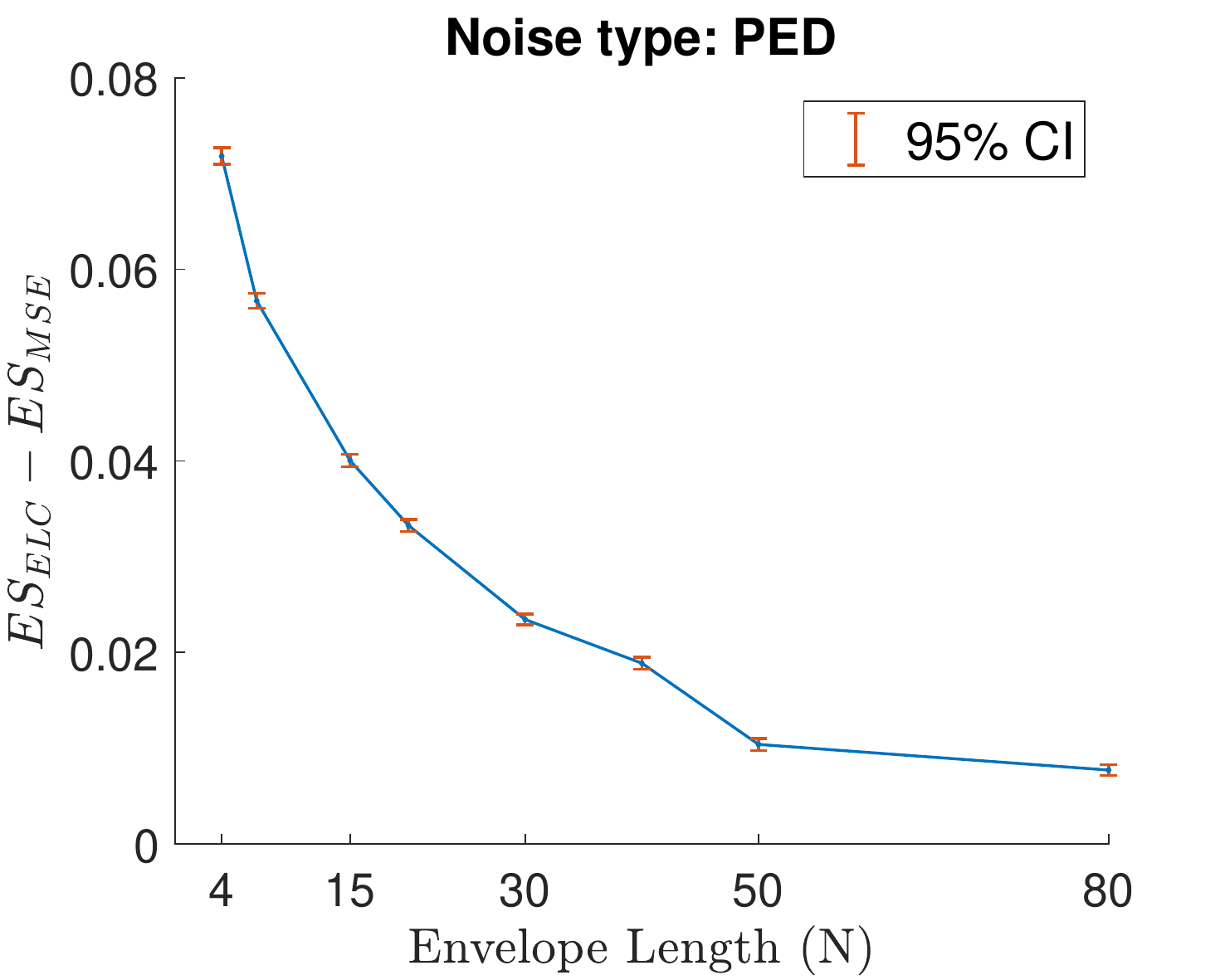}}
		\label{fig:test6}
	\end{minipage}
	\caption{Average ELC differences, as function of envelope durations $N$, between $\text{ES}_{ELC}$ and $\text{ES}_{MSE}$ systems, for different noise types. We observe a monotonic decreasing relationship between the average ELC difference and the envelope length and for $N=80$, the average ELC difference between the $\text{ES}_{ELC}$ and $\text{ES}_{MSE}$ systems is close to zero. This is in line with the theoretical results of Sec.\;\ref{secrelation}.}
	\label{fig:pernoisetype}
\end{figure*}
\subsection{Comparing One-third Octave Bands}
In Fig.\;\ref{fig:perband} we present the ELC scores, as function of envelope duration $N$, for each of the $J=15$ one-third octave band DNNs in the $\text{ES}_{ELC}$ and $\text{ES}_{MSE}$ systems. All DNNs are tested using speech corrupted with BBL noise at an SNR of 0 dB. 
First, we observe that both systems manage to improve the ELC score considerably, when compared to the ELC score of the noisy speech signals, i.e. both systems enhance the noisy speech, which is in line with known results \cite{kolbaek_speech_2017}. 
Furthermore, we can observe that the DNNs trained with the ELC cost function, i.e. the $\text{ES}_{ELC}$ systems, in general achieve higher, or similar, ELC scores than the DNNs trained with the STSA-MSE cost function, i.e. the $\text{ES}_{MSE}$ systems. 
This is an important observation, since it verifies that DNNs trained to maximize ELC indeed achieve the highest, or similar, ELC scores compared to DNNs trained to optimize a different cost function, STSA-MSE in this case.    
Finally, and most importantly, we observe that the difference in ELC score between the $\text{ES}_{ELC}$ and $\text{ES}_{MSE}$ DNNs generally decrease with increasing $N$. For $N=80$ the ELC score of the $\text{ES}_{ELC}$ and $\text{ES}_{MSE}$ DNNs practically coincide. 
\subsection{Comparing ELC across Noise Types}
In Fig.\;\ref{fig:pernoisetype} we present the ELC score difference, as function of envelope duration $N$, for $\text{ES}_{ELC}$ and $\text{ES}_{MSE}$ systems, when tested using speech material corrupted with various noise types at an SNR of 0 dB. 
Specifically, we compute the difference in ELC score for each pair of one-third octave band DNNs in the $\text{ES}_{ELC}$ and $\text{ES}_{MSE}$ systems, and then compute the average ELC difference as function of envelope duration $N$. We do this for all the 1000 test utterances and for each of the six noise types introduced in Sec\;\ref{sec:nt}: SSN, BBL, CAF, STR, BUS, and PED. 
Finally, we compute the $95\%$ confidence interval\,(CI) on the mean ELC difference.

From Fig.\;\ref{fig:pernoisetype} we observe that the average ELC difference, i.e. $\text{ES}_{ELC} - \text{ES}_{MSE}$, appears to be monotonically decreasing with respect to the duration of the envelope $N$. 
Furthermore, we observe that the average ELC difference approaches zero as the duration of the envelope $N$ increases, and similarly to Fig.\;\ref{fig:perband}, for $N=80$, the difference between the $\text{ES}_{ELC}$ and $\text{ES}_{MSE}$ systems is close to zero.  
Finally, we observe that the $95\%$ confidence intervals are relatively narrow for all envelope durations and noise types, which indicate that our test set is sufficiently large to provide accurate estimates of the true mean ELC difference.        
Similarly to Fig.\;\ref{fig:perband}, the results in Fig.\;\ref{fig:pernoisetype} support the theoretical results of Sec.\;\ref{secrelation}. 
Additionally, the results in Fig.\;\ref{fig:pernoisetype} show consistency across multiple noise types, which suggests that the theory in practice applies for various noise type distributions.

{\color{black}
\subsection{Comparing STOI across Noise Types}
We now investigate if the global behavior observed for approximate-STOI, i.e. ELC, in Fig.\;\ref{fig:pernoisetype} also applies for real STOI.  
To do this, we reconstruct the test signals used for Fig.\;\ref{fig:pernoisetype} in the time domain. We follow the technique proposed in \cite{kolbaek_monaural_2018-1}, where a uniform gain across STFT coefficients within a one-third octave band is used before an inverse DFT is applied using the phase of the noisy signal.   
In Table\,\ref{tab:stoi_scorr1} we present the STOI scores for $\text{ES}_{ELC}$ and $\text{ES}_{MSE}$ systems, as a function of $N$, when tested using speech material corrupted with different noise types at an SNR of 0 dB.
Note that these test signals are similar to the test signals used for Fig.\;\ref{fig:pernoisetype} except that we now evaluate them according to STOI and not ELC. 

From Table\,\ref{tab:stoi_scorr1} we observe that the average STOI difference between the $\text{ES}_{ELC}$ and $\text{ES}_{MSE}$ systems is maximum for $N=4$, but quickly tends to zero as $N$ increases and for $N\ge15$, the STOI difference is practically zero, i.e. $\leq 0.01$. 
{\color{black}
Also, we observe that the gap in STOI between the $\text{ES}_{ELC}$ and $\text{ES}_{MSE}$ systems closes faster at a lower value of $N$ in Table\;\ref{tab:stoi_scorr1} compared to Fig.\;\ref{fig:pernoisetype}.
We believe this is due to the transformation of the, potentially "invalid", sequences of (e.g. \cite{nawab_signal_1983,griffin_signal_1984}) modified magnitude spectra, when reconstructing enhanced time-domain signals, whose intelligibility is estimated by STOI in  Table\;\ref{tab:stoi_scorr1}. 
Therefore, STOI in Table\,\ref{tab:stoi_scorr1} might be computed based on slightly different magnitude spectra compared to the magnitude spectra used for computing the ELC scores in Fig.\;\ref{fig:pernoisetype}. 
Furthermore, we observe that the $\text{ES}_{MSE}$ achieve slightly higher STOI scores than the $\text{ES}_{ELC}$ systems for $N=4$, which might be due to sub-optimal learning rates as the ones actually used during training of the systems at, e.g. $N=4$,  were found based on a grid-search using systems with $N=30$ (see Sec. VI.D). More importantly, the maximum improvement in STOI is achieved for $N=\{15,20,30\}$, where both systems achieve similar STOI scores.} 
Finally, while the theoretical results of Sec.\;\ref{secrelation} show that approximate-STOI performance of $\ah_{MMELC}$ and $\ah_{MMSE}$ is identical, asymptotically, for $N \to \infty$, the empirical results in Table\,\ref{tab:stoi_scorr1} suggest that $N\ge15$ is sufficient for practical equality to hold for DNN based speech enhancement systems. 
\begin{table}
	\caption{{\color{black} STOI scores as function of $N$ for $\text{ES}_{ELC}$ and $\text{ES}_{MSE}$ systems tested using different noise types at an SNR of 0 dB.}}
	\label{tab:stoi_scorr1}
	\centering
	\setlength\tabcolsep{5pt} 
	\resizebox{1.0\columnwidth}{!}{%
		{\color{black}\begin{tabular}{cccccccccc}
				\toprule
				$N:$ & & $4$ & $7$ & $15$ & $20$ & $30$ & $40$ & $50$ & $80$ \\
				\midrule
				\multirow{ 2}{*}{SSN:} & ELC :& 0.81 & 0.85 & 0.88 & 0.88 & 0.87 & 0.86 & 0.85 & 0.84 \\
				& MSE :& 0.84 & 0.87 & 0.87 & 0.87 & 0.87 & 0.86 & 0.85 & 0.84 \\ [6pt]
				\multirow{ 2}{*}{BBL:} & ELC :& 0.77 & 0.80 & 0.82 & 0.82 & 0.81 & 0.80 & 0.80 & 0.78 \\
				& MSE :& 0.79 & 0.82 & 0.82 & 0.82 & 0.81 & 0.80 & 0.80 & 0.78 \\ [6pt]
				\multirow{ 2}{*}{CAF:} & ELC :& 0.82 & 0.85 & 0.87 & 0.87 & 0.86 & 0.85 & 0.84 & 0.83 \\
				& MSE :& 0.85 & 0.87 & 0.87 & 0.87 & 0.86 & 0.85 & 0.85 & 0.84 \\ [6pt]
				\multirow{ 2}{*}{STR:} & ELC :& 0.83 & 0.86 & 0.88 & 0.89 & 0.88 & 0.87 & 0.87 & 0.85 \\
				& MSE :& 0.86 & 0.88 & 0.88 & 0.88 & 0.88 & 0.87 & 0.87 & 0.85 \\ [6pt]
				\multirow{ 2}{*}{PED:} & ELC :& 0.77 & 0.81 & 0.83 & 0.83 & 0.83 & 0.82 & 0.81 & 0.80 \\
				& MSE :& 0.80 & 0.82 & 0.83 & 0.83 & 0.82 & 0.82 & 0.81 & 0.80 \\ [6pt]
				\multirow{ 2}{*}{BUS:} & ELC :& 0.87 & 0.89 & 0.90 & 0.91 & 0.90 & 0.89 & 0.89 & 0.89 \\
				& MSE :& 0.89 & 0.90 & 0.90 & 0.90 & 0.90 & 0.90 & 0.89 & 0.89 \\ [3pt] \bottomrule 
	\end{tabular}}}
\end{table}

\subsection{Comparing Gain-Values}
{\color{black}
Figures\;\ref{fig:perband} and \ref{fig:pernoisetype}, and Table\,\ref{tab:stoi_scorr1} show that $\text{ES}_{ELC}$ systems achieve approximately the same ELC and STOI values as $\text{ES}_{MSE}$ systems and that the ELC and STOI difference between the two types of systems approach zero as $N$ becomes large. These empirical results are in line with the theoretical results in Sec.\;\ref{secrelation}. }
However, the results in Sec.\;\ref{secrelation} predict that not only do $\text{ES}_{ELC}$, and $\text{ES}_{MSE}$ systems produce identical ELC scores, they also predict that the systems are, in fact, essentially identical, i.e. up to an affine transformation. Hence, in this section, we compare how the systems actually operate. 
Specifically, we compare the gains estimated by $\text{ES}_{ELC}$ systems with gains estimated by $\text{ES}_{MSE}$ systems.      

In Fig.\;\ref{fig:gaincorr} we present scatter plots, one for each one-third octave band for pairs of gains estimated by $\text{ES}_{ELC}$ and $\text{ES}_{MSE}$ systems tested with BBL noise at an SNR of 5 dB. Each scatter plot consists of 10000 pairs of gains acquired by sampling 10 gain-pairs randomly and uniformly distributed from each of the 1000 test utterances. In Fig.\;\ref{fig:gaincorr}, yellow indicates high density of gain-pairs and dark blue indicates low density.   
{\color{black}
From Fig.\;\ref{fig:gaincorr} it is seen that a correlation no smaller than $0.88$ is achieved for all 15 one-third octave bands. The highest correlation of $r=0.98$ is achieved by bands $5$ to $7$ and the lowest is $r=0.88$ achieved by band $2$ followed by band $1$ with $r=0.89$. 
}
It is also seen that a large number of gain values are either zero, or one, as one would expect due to the sparse nature of speech in the T-F domain. 
However, although a strong correlation is observed for all bands, the gain-pairs are slightly more scattered at the first few bands than for the remaining bands. 
This might be explained simply by the fact that low one-third octave bands correspond to single STFT bins, whereas higher one-third octave bands are sums of a large number of STFT bins. 
This, in turn, may have the consequence that for finite $N$ $(N=30)$, Kolmogorovs strong law of large numbers (see Appendix.\;\ref{sec:zindepedent}) is better valid at higher frequencies than at lower frequencies (so that gain vectors produced by one system is closer to an affine transformation of gain vectors produced by the other system).
In fact, if we compute $r_1$ for models trained with $N=50$, we get $r_1 = 0.93$, i.e. increased correlation between the gain vectors produced by the two systems.  
Finally, in Table.\;\ref{tab:lincorr} we present average correlation coefficients and we observe correlation coefficients  $\geq 0.87$ for all, both matched and unmatched, noise types, at multiple SNRs.
%
%
\begin{figure*}[ht] 
	\centering
	\centerline{\includegraphics[trim={20mm 6mm 20mm 6mm},clip,width=1.0\linewidth]{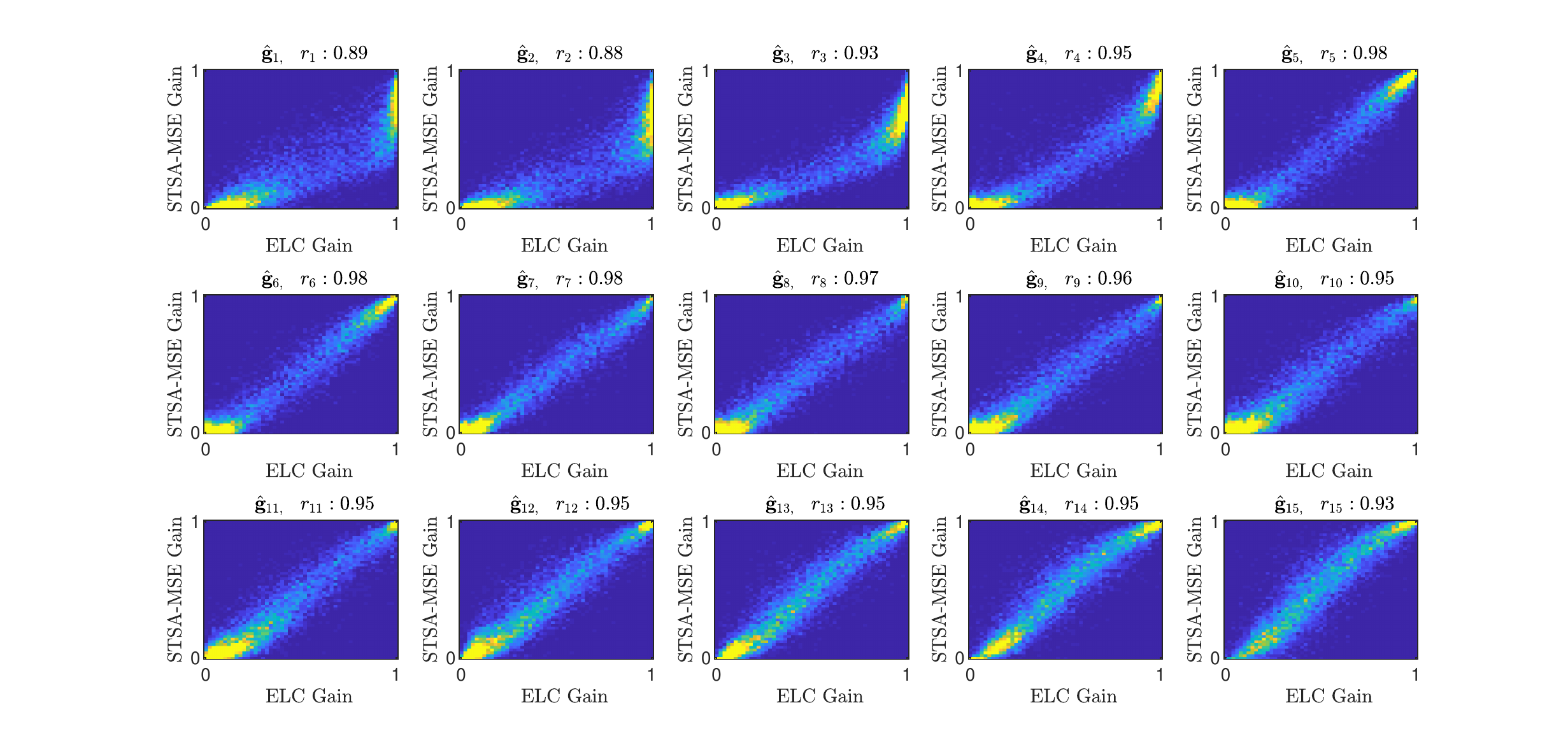}}
	\caption{{\color{black}Scatter plots based on gain values from $\text{ES}_{ELC}$ and $\text{ES}_{MSE}$ systems with an envelope length of $N=30$. Dark blue indicate low density and bright yellow indicate high density. The systems are tested with BBL noise corrupted speech at an SNR of 5 dB. Each figure shows one of 15 ($\hat{\mathbf{g}}_1, \hat{\mathbf{g}}_2, \dots, \hat{\mathbf{g}}_{15}$) one-third octave bands. A correlation no smaller than $0.88$ is achieved for all one-third octave bands, which indicates that the $\text{ES}_{ELC}$ and $\text{ES}_{MSE}$ systems estimate fairly similar gain vectors. }}
	\label{fig:gaincorr}
\end{figure*}
\begin{table}
\caption{Sample correlations between gains from $\text{ES}_{ELC}$ and $\text{ES}_{MSE}$ systems with $N=30$. See Fig.\;\ref{fig:gaincorr} for per band correlations.}
\label{tab:lincorr}
\centering
\setlength\tabcolsep{5pt} 
\resizebox{0.8\columnwidth}{!}{%
	\begin{tabular}{ccccccc}
		\toprule
		\begin{tabular}[c]{@{}c@{}}SNR \\ {[dB]}\end{tabular} 	& 
		\begin{tabular}[c]{@{}c@{}}SSN \\\end{tabular} 	& 
		\begin{tabular}[c]{@{}c@{}}BBL \\\end{tabular} 	& 
		\begin{tabular}[c]{@{}c@{}}CAF \\\end{tabular} 	&
		\begin{tabular}[c]{@{}c@{}}STR \\\end{tabular} 	&
		\begin{tabular}[c]{@{}c@{}}BUS \\\end{tabular} 	&
		\begin{tabular}[c]{@{}c@{}}PED \\\end{tabular} 	\\
		\midrule
		-5   & 0.94 & 0.87 & 0.89 & 0.93  & 0.87 & 0.90\\ 
		0    & 0.94 & 0.92 & 0.92 & 0.93  & 0.88 & 0.92\\ 
		5    & 0.95 & 0.95 & 0.93 & 0.93  & 0.90 & 0.92\\ 
		10   & 0.95 & 0.95 & 0.92 & 0.92  & 0.91 & 0.93\\ \bottomrule
\end{tabular}}
\end{table}

\section{Conclusion}\label{sec:con}
This study is motivated by the fact that most estimators used for speech enhancement, being either data-driven models, e.g. deep neural networks\,(DNNs), or statistical model-based techniques such as the short-time spectral amplitude minimum mean-square error\,(STSA-MMSE) estimator, use the STSA mean-square error\,(MSE) cost function as a performance indicator.  
Short-time objective intelligibility\,(STOI), a state-of-the-art speech intelligibility estimator, on the other hand, rely on the envelope linear correlation\,(ELC) of speech temporal envelopes. 
Since the primary goal of many speech enhancement systems is to improve speech intelligibility, it raises the question if estimators can benefit from an ELC cost function. 

In this paper we derive the maximum mean envelope linear correlation\,(MMELC) estimator and study its relationship to the well-known STSA-MMSE estimator.    
We show theoretically that the MMELC estimator, under a commonly used conditional independence assumption, is asymptotically equivalent to the STSA-MMSE estimator.
Furthermore, we demonstrate experimentally that this relationship also holds for DNN based speech enhancement systems, when the DNNs are trained to either maximize ELC or minimize MSE and the systems are evaluated using both ELC and STOI.
Finally, our experimental findings suggest, that applying the traditional STSA-MMSE estimator on noisy speech signals in practice leads to essentially maximum speech intelligibility as reflected by the STOI speech intelligibility estimator.	 
%
%


%

\appendices

\section{Maximizing a Constrained Inner Product}\label{sec:lagran}
This appendix derives an expression for the zero-mean, unit-norm vector $\e(\ah)$, which maximizes the inner product with the vector $\E_{\A | \sr } \left[ \e(\A| \sr) \right]$.
For notational convenience, let $\alfa = \E_{\A | \sr } \left[ \e(\A| \sr) \right]$, and $\bet = \e(\ah)$.  The constrained optimization problem from Eq.\;\eqref{eq6} is then defined as  
\begin{equation}
\begin{aligned}
& \underset{\bet}{\text{maximize}}
& & \alfa^T \bet \\
& \text{subject to} 
& & \bet^T \One = 0, \\
& & &\bet^T\bet = 1.
\end{aligned}
\label{eq:eqLag1}
\end{equation}
The vector $\bet^\ast$ that solves Eq.\;\eqref{eq:eqLag1} can be found using the method of Lagrange multipliers \cite{boyd_convex_2004}. Introducing two scalar Lagrange multipliers, $\lambda_1$ and $\lambda_2$, for the two equality constraints, the Lagrangian is given by\footnote{We solve the equivalent problem that minimizes $-\alfa^T \bet$.}    
\begin{equation}
\mathcal{L}(\bet,\lambda_1,\lambda_2) = -\alfa^T \bet + \lambda_1 \bet^T \One + \lambda_2(\bet^T\bet - 1).
\label{eqLag2}
\end{equation}
Setting the partial derivatives $\frac{\partial \mathcal{L}}{\partial \bet}$ equal to zero 
\begin{equation}
\begin{aligned}
\frac{\partial \mathcal{L}}{\partial \bet} & = -\alfa + \lambda_1 \One + 2\lambda_2\bet = \barbelow{0}, \\
\end{aligned}
\label{eqLag30}
\end{equation}
and solving for $\bet$, we arrive at 
\begin{equation}
\begin{aligned}
\bet & = \frac{\alfa - \lambda_1 \One}{2\lambda_2}.
\end{aligned}
\label{eqLag3}
\end{equation}
Using the same approach for $\frac{\partial \mathcal{L}}{\partial \lambda_1}$ and $\frac{\partial \mathcal{L}}{\partial \lambda_2}$, substituting in Eq.\;\eqref{eqLag3} and solving for $\lambda_1$, and $\lambda_2$ such that the two constraints are fulfilled, we find   
\begin{equation}
\begin{aligned}
\lambda_1 & =   \frac{1}{N} \alfa^T \One =  {\mu}_{\alfa},\\
\end{aligned}
\label{eqLag40}
\end{equation}
and
\begin{equation}
\begin{aligned}
\lambda_2 & =  \frac{\lVert \alfa - \barbelow{\mu}_{\alfa} \One \rVert }{2}.
\end{aligned}
\label{eqLag4}
\end{equation}
Inserting $\lambda_1$ and $\lambda_2$ into Eq.\;\eqref{eqLag3} results in 
\begin{equation}
\begin{aligned}
\bet^\ast & = \frac{ \alfa - \barbelow{\mu}_{\alfa} \One  }{\lVert \alfa - \barbelow{\mu}_{\alfa} \One \rVert},
\end{aligned}
\label{eqLag5}
\end{equation}
which is simply the vector $\alfa$, normalized to zero sample mean and unit norm.

\section{Factorization of Expectation}\label{sec:zindepedent}
This appendix shows that the expectation in Eq.\;\eqref{eq11} factorizes into the product of expectations in Eq.\;\eqref{eq14},  asymptotically as $N \to \infty$. 
Let
\begin{equation} 
\Y \triangleq \A | \sr,
\label{eqz1}
\end{equation}
and
\begin{equation} 
\Hidem \triangleq  \barbelow{\barbelow{I}}_N - \frac{1}{N} \One\One^T,
\label{eqz2}
\end{equation}
so that
\begin{equation} 
\Z = \Hidem \Y,
\label{eqz3}
\end{equation}
where $\barbelow{\barbelow{I}}_N$ denotes the $N$-dimensional identity matrix and $\A | \sr $ is a random vector distributed according to the conditional probability density function $\fagr$.
A specific element $Z_i$, of $\Z$ is then given by
\begin{equation} 
\begin{split}
Z_i &= \barbelow{{h}}_i^T \Y \\
& =   S_i - \frac{1}{N} \One^T \Y,  \\
\end{split}
\label{eqz4}
\end{equation} 
where $\barbelow{{h}}_i$ is the $i$th column of matrix $\Hidem$. 

We now define the covariance between $Z_i$ and $1/\lVert \Z \rVert$ as
\begin{equation} 
\begin{split}
\text{cov}(Z_i , \frac{1}{\big\lVert \Z \big\rVert} ) &\triangleq \E \left[ \bigg( Z_i - \E \left[ Z_i \right] \bigg) \bigg( \frac{1}{ \big\lVert \Z \big\rVert } - \E \left[ \frac{1}{ \big\lVert \Z \big\rVert } \right] \bigg)    \right]  \\
& = \E \left[ \frac{  Z_i }{ \big\lVert \Z \big\rVert }  \right] - \E \left[ Z_i \right] \E \left[ \frac{1}{\big\lVert \Z \big\rVert } \right].  \\
\end{split}
\label{eqz5}
\end{equation} 
We can rewrite the factors on the right-hand side of Eq.\;\eqref{eqz5} as follows
\begin{equation} 
\begin{split}
\E \left[ Z_i \right] &= \E \left[ \barbelow{\mathbf{h}}_i^T \Y  \right]  \\
& = \E \left[ S_i - \frac{1}{N} \One^T \Y  \right]  \\
& = \E \left[ S_i  \right] - \frac{1}{N} \One^T  \E \left[ \Y  \right]  \\
& = \E \left[ S_i  \right] - \frac{1}{N} \sum_{j=1}^{N}  \E \left[ S_j  \right],  \\
\end{split}
\label{eqz6}
\end{equation} 
\begin{equation} 
\begin{split}
\E \left[ \frac{1}{\big\lVert \Z \big\rVert } \right] &= \E \left[ \frac{1}{ \sqrt{ \Y^T \Hidem \Hidem^T \Y} }  \right]   \\
& =  \E \left[ \frac{1}{ \sqrt{ \Y^T \Hidem \Y} }  \right]   \\
& =  \E \left[ \frac{1}{ \sqrt{ \Y^T\Y  - \frac{1}{N} \Y^T \One \One^T \Y} }  \right]   \\
& =  \E \left[ \frac{1}{ \sqrt{ \sum_{j=1}^{N}S_j^2  - \frac{1}{N} \left( \sum_{j=1}^{N}S_j \right)^2 } }  \right]   \\
& =  \E \left[ \frac{\sqrt{\frac{1}{N}}}{ \sqrt{ \frac{1}{N} \sum_{j=1}^{N}S_j^2  - \left( \frac{1}{N}  \sum_{j=1}^{N}S_j \right)^2 } }  \right],   \\
\end{split}
\label{eqz7}
\end{equation} 
and
\begin{equation} 
\begin{split}
\E \left[ \frac{  Z_i }{ \big\lVert \Z \big\rVert }  \right] &= 	\E \left[ \frac{  \sqrt{ \frac{1}{N}}  \left( S_i - \frac{1}{N} \sum_{j=1}^{N}S_j \right)}{ \sqrt{ \frac{1}{N} \sum_{j=1}^{N}S_j^2  - \left( \frac{1}{N} \sum_{j=1}^{N}S_j \right)^2 } }  \right].   \\
\end{split}
\label{eqz8}
\end{equation} 

In Eqs.\;\eqref{eqz6}, \eqref{eqz7} and \eqref{eqz8} two different sums of random variables occur, 
\begin{equation} 
\begin{split}
\frac{1}{N}  \sum_{j=1}^{N}S_j, \\
\end{split}
\label{eqz10}
\end{equation}     
and
\begin{equation} 
\begin{split}
\frac{1}{N} \sum_{j=1}^{N}S_j^2.  \\
\end{split}
\label{eqz11}
\end{equation} 
Since, by assumption, Eq.\;\eqref{eqz9}, $S_j\;\forall \;j$ are independent random variables with finite variances\footnote{Assuming a finite variance of $S_j$ is motivated by the fact that $S_j$ model speech signals, which always take finite values due to both physical and physiological limitations of sound and speech production systems, respectively.}%
, according to Kolmogorovs strong law of large numbers \cite{sen_large_1994}, the sums given by Eqs.\;\eqref{eqz10} and \eqref{eqz11} will converge (almost surely, i.e. with probability\;($\Pr$) one) to their average means $\mu_S = \frac{1}{N}\sum_{j=1}^{N}\E[S_j]$, and $\mu_{S^2} = \frac{1}{N}\sum_{j=1}^{N}\E[S_j^2]$, respectively, as $N \to \infty$. Formally, we can express this as  
\begin{equation} 
\begin{split}
\Pr &\left( \lim\limits_{N \to \infty} \frac{1}{N}  \sum_{j=1}^{N}S_j = \mu_{S} \right) = 1, \\
\end{split}
\label{eqz12}
\end{equation}     
and
\begin{equation} 
\begin{split}%
\Pr &\left( \lim\limits_{N \to \infty} \frac{1}{N}  \sum_{j=1}^{N}S_j^2 = \mu_{S^2} \right) = 1. \\
\end{split}
\label{eqz13}
\end{equation}     

By substituting Eqs.\;\eqref{eqz12}, and \eqref{eqz13} into Eqs.\;\eqref{eqz6}, \eqref{eqz7} and \eqref{eqz8}, we arrive at
\begin{equation} 
\begin{split}
\lim\limits_{N \to \infty} \E \left[ Z_i \right] &= \E \left[ S_i  \right] - \mu_S,  \\
\end{split}
\label{eqz14}
\end{equation} 
\begin{equation} 
\begin{split}
\lim\limits_{N \to \infty} \E \left[ \frac{1}{\big\lVert \Z \big\rVert } \right] &= \frac{\lim\limits_{N \to \infty} \sqrt{\frac{1}{N}}}{ \sqrt{ \mu_{S^2}  - \mu_S^2 } }     ,    \\
\end{split}
\label{eqz15}
\end{equation} 
and
\begin{equation} 
\begin{split}
\lim\limits_{N \to \infty} \E \left[ \frac{  Z_i }{ \big\lVert \Z \big\rVert }  \right] &=   \left(\E \left[ S_i  \right] - \mu_S \right) \frac{\lim\limits_{N \to \infty} \sqrt{\frac{1}{N}}  }{ \sqrt{ \mu_{S^2}  - \mu_S^2 } }  \\
& = \lim\limits_{N \to \infty} \E \left[ Z_i  \right]   \E \left[ \frac{  1 }{ \big\lVert \Z \big\rVert }  \right]  ,   \\
\end{split}
\label{eqz16}
\end{equation} 
where the last line follows from Eq.\;\eqref{eqz14} and \eqref{eqz15}.
In words, as $N \to \infty$, the covariance between $Z_i$ and $1/\lVert \Z \rVert$ tends to zero and, consequently, the expectation in Eq.\;\eqref{eq11} factorizes into the product of expectations in Eq.\;\eqref{eq14}.


%
%
%

\bibliographystyle{bib/IEEEtran}
\bibliography{bib/mybib}

%

\vspace{-10mm}
\begin{IEEEbiography}[{\includegraphics[width=1in,height=1.25in,clip,keepaspectratio]{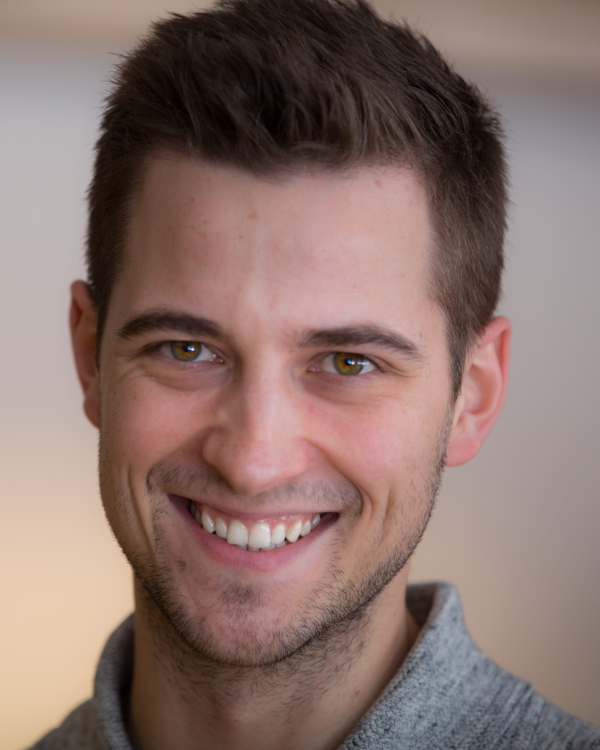}}]%
	{Morten Kolbæk}
	received the B.Eng. degree in electronic design at Aarhus University, Business and Social Sciences, AU Herning, Denmark, in 2013 and the M.Sc. in signal processing and computing from Aalborg University, Denmark, in 2015. 
	He is currently pursuing his PhD degree at the section for Signal and Information Processing at the Department of Electronic Systems, Aalborg University, Denmark.
	His research interests include speech enhancement and separation, deep learning, and intelligibility improvement of noisy speech. 
\end{IEEEbiography}
\vspace{-10mm}
\begin{IEEEbiography}
	[{\includegraphics[width=1in,height=1.25in,clip,keepaspectratio]{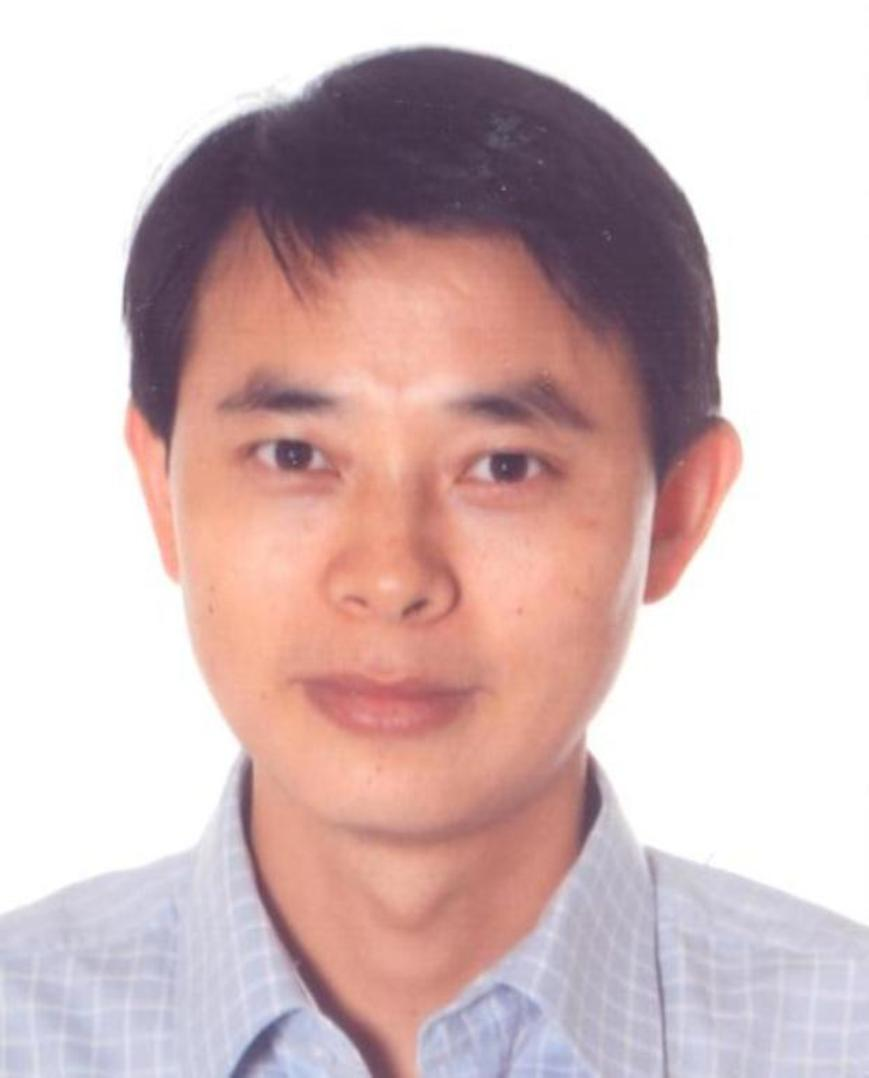}}]
	{Zheng-Hua Tan}
	(M'00--SM'06) received the B.Sc. and M.Sc. degrees in electrical engineering from Hunan University, Changsha, China, in 1990 and 1996, respectively, and the Ph.D. degree in electronic engineering from Shanghai Jiao Tong University, Shanghai, China, in 1999. 
	He is a Professor and a Co-Head of the Centre for Acoustic Signal Processing Research (CASPR) at Aalborg University, Aalborg, Denmark. He was a Visiting Scientist at the Computer Science and Artificial Intelligence Laboratory, MIT, Cambridge, USA, an Associate Professor at Shanghai Jiao Tong University, and a postdoctoral fellow at KAIST, Daejeon, Korea. His research interests include machine learning, deep learning, pattern recognition, speech and speaker recognition, noise-robust speech processing, multimodal signal processing, and social robotics. He is a member of the IEEE Signal Processing Society Machine Learning for Signal Processing Technical Committee (MLSP TC). He is an Editorial Board Member for Computer Speech and Language and was a Guest Editor for the IEEE Journal of Selected Topics in Signal Processing and Neurocomputing. He was the General Chair for IEEE MLSP 2018 and a TPC co-chair for IEEE SLT 2016.
\end{IEEEbiography}
\vspace{-10mm}
\begin{IEEEbiography}[{\includegraphics[width=1in,
	height=1.25in,clip,keepaspectratio]{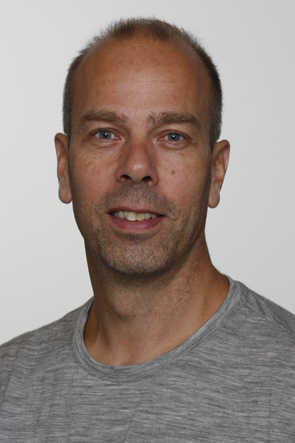}}]{Jesper Jensen} received the M.Sc. degree in electrical engineering and the Ph.D. degree in signal processing from Aalborg University, Aalborg, Denmark, in 1996 and 2000, respectively. From 1996 to 2000, he was with the Center for Person Kommunikation (CPK), Aalborg University, as a Ph.D. student and Assistant Research Professor. From 2000 to 2007, he was a Post-Doctoral Researcher and Assistant Professor with Delft University of Technology, Delft, The Netherlands, and an External Associate Professor with Aalborg University. Currently, he is a Senior Principal Scientist with Oticon A/S, Copenhagen, Denmark, where his main responsibility is scouting and development of new signal processing concepts for hearing aid applications. He is a Professor with the Section for Signal and Information Processing (SIP), Department of Electronic Systems, at Aalborg University. He is also a co-founder of the Centre for Acoustic Signal Processing Research (CASPR) at Aalborg University. His main interests are in the area of acoustic signal processing, including signal retrieval from noisy observations, coding, speech and audio modification and synthesis, intelligibility enhancement of speech signals, signal processing for hearing aid applications, and perceptual aspects of signal processing.
\end{IEEEbiography}

\end{document}